\documentstyle[psfig,preprint,aps]{revtex}
\begin{document}

\draft
\title{High Density QCD }
\author{
 M. B. Gay Ducati
$^{\star}$\footnotetext{$^{\star}$E-mail:gay@if.ufrgs.br} }  
\address{ Instituto de F\'{\i}sica, Universidade
Federal do Rio Grande do Sul\\ Caixa Postal 15051, CEP 91501-970, Porto
Alegre, RS, BRAZIL}

\maketitle

\begin{abstract}
{\small The dynamics of high partonic density QCD is presented considering, in
the double logarithm approximation, the parton recombination mechanism built in
the AGL formalism, developed including unitarity corrections for the  nucleon
as well for nucleus.  It is shown that these corrections are under theoretical
control. The resulting non linear evolution equation is solved in the
asymptotic regime, and a comprehensive phenomenology concerning Deep Inelastic
Scattering like $F_2$, $F_L$, $F_2^c$. $\partial F_2/ \partial \ln Q^2$,
$\partial F^A_2/ \partial \ln Q^2$, etc, is presented.
 
The connection of our formalism with the DGLAP and BFKL dynamics, and with
other perturbative (K) and non-perturbative (MV-JKLW) approaches
is analised in detail. The phenomena of saturation due to
shadowing corrections and the relevance of this effect in ion
physics and heavy quark production is emphasized. The
implications to $e$-RHIC, HERA-A, and LHC physics and some open
questions are mentioned.}

\end{abstract}

\vspace{2cm}

{\Large  Plenary Talk presented at XXI ENFPC, S\~ao Louren\c{c}o, Brasil,
October, 24$^{\rm{th}}$ (2000).}

\vspace{3cm}
\bigskip

\section{Introduction}

The dynamics of the high density Quantum Chromodynamics (hdQCD) is one of
the present most challeging open questions in high energy physics. The intense
theoretical and experimental activity towards the understanding of small $x$
(small fraction of proton  momentum carried by the struck parton) QCD takes
place from Deep Inelastic Scattering (DIS) at HERA \cite{DIS} to heavy ions
collisions (HIC) at RHIC \cite{RHIC}. This kinematical regime will also be
tested at LHC in a near future \cite{LHC}.

Important contribution to the
interest of the field is due to the puzzling result obtained by HERA at
small-$x$ ($x\leq 10^{-2}$) \cite{r1} for the proton structure function
$F_2(x,Q^2)$. This function was observed to increase dramatically as $x$ gets
smaller (Fig. 1). In the region of moderate Bjorken $x$ ($x\geq 10^{-2}$) the
Operator Product Expansion (OPE) methods as well as the Renormalization Group
Equations (RGE) have been applied successfully \cite{r2}. The evolution of
quark and gluon distribution functions given by the DGLAP equations
\cite{r3} is based on the summing of the leading powers of $\alpha_s \ln Q^2
\approx 1$, $\alpha_s \ln (1/x) <<1$, $\alpha_s <<1$,  where $\alpha_s$ is the
strong coupling constant. The leading $\ln (1/x)$ contributions is  the case
for the BFKL equation \cite{BFKL}.  The  procedure  known as the double leading
logarithmic approximation (DLA) corresponds in axial gauges to generate the
logarithms by ladder diagrams, whose emitted gluons have strongly ordered  
transverse and longitudinal momenta, summing the logs $\alpha_s \ln Q^2 \ln
(1/x)$. It was shown that the DLA is a common
 limit between the linear
dynamics \cite{r4}.

 The increasing of the parton densities requires a formulation of
the QCD at high partonic density, where unitarity corrections (UC), not
considered in the previous dynamics already mentioned, are properly taken into
account. In this sense, the small $x$ region, where the gluon distribution
sets the behavior of the main observables, provides the interface between
perturbative and non perturbative QCD, or in other words, between hard and
soft physics. Clearly, both experimentalists and theoreticians are challenged
to desentangle, measure and formulate the dynamical collective effects that
are subjacent to the observed result of increasing $F_2$  and the cross
section $\sigma_{tot}$ at DIS,  as $x$ gets smaller \cite{r1}. Both evolution
equations, DGLAP (evolution in $\ln Q^2$) and BFKL (evolution in
$\ln(1/x)$) as representatives of linear dynamics, need control in order to
restore unitarity, since the Froissart limit requires $\sigma_{tot} \leq Cte
\,\ln^2 s$ \cite{r7}. 

A comprehensive
treatment should envolve both linear and non linear regimes. The main attempts
to develop a formalism for hdQCD are the approaches of Mc.Lerran and
collaborators ($MVJKLW$) \cite{McLerran}, by Kovchegov ($K$) \cite{Kovchegov}
and by Ayala, Gay Ducati and Levin ($AGL$) \cite{AGL,AGL1}. Derived
independently, the three methods obtain non linear evolution equations for the
gluon distribution, at the small $x$ region, describing the onset of hdQCD,
although considering different degrees of freedom.

In what follows I will present an introductory review of the subject of hdQCD,
the main aspects of the formulations to the subject, the connections among
them in the asymptotic region ($x \rightarrow 0$), present the state of art of
the phenomenology of small $x$ physics and address some open questions. 

\section{The Evolution Equations} 

It will be briefly presented
the DGLAP and BFKL dynamics and their predictions for the small $x$
regime at DIS. The DIS is the process of interaction of a lepton and a nucleon
exchanging an electroweak boson producing many particles at the final state,
which is a hadronic state $X$. The process is 
\begin{eqnarray}
l(k) \, \, N(p) \rightarrow l^{\prime} (k^{\prime}) \,\, X \,\,,
\end{eqnarray} where
$k$, $k^{\prime}$, $p$ and $p^{\prime}$ are the fourmomenta of the  initial and
final lepton, incident nucleon and final hadronic system, respectively. The
main variables for this process are $Q^2=-q^2= -(k-k^{\prime})^2$, which is the
square of the transfered momentum, $s=(k+p)^2$, the square of the
center of mass energy, $W^2=(q+p)^2=(p^{\prime})^2$, the square of the center
of mass energy of the virtual boson-nucleon  system. The hard scale is given by
$q^2$ ($<0$), corresponding to the process resolution, and $x=Q^2/2p.q$,
meaning the virtual boson resolves the hadronic structure, or the partons, 
since $\Delta x \approx 1/ \sqrt{Q^2}$ . Are also useful the variables
$ y = q.p / k.p $, measuring the process inelasticity and $ \nu = q.p/m_N$, the energy
of the virtual boson once taken in the target rest frame.
 
\begin{figure}
\centerline{\psfig{file=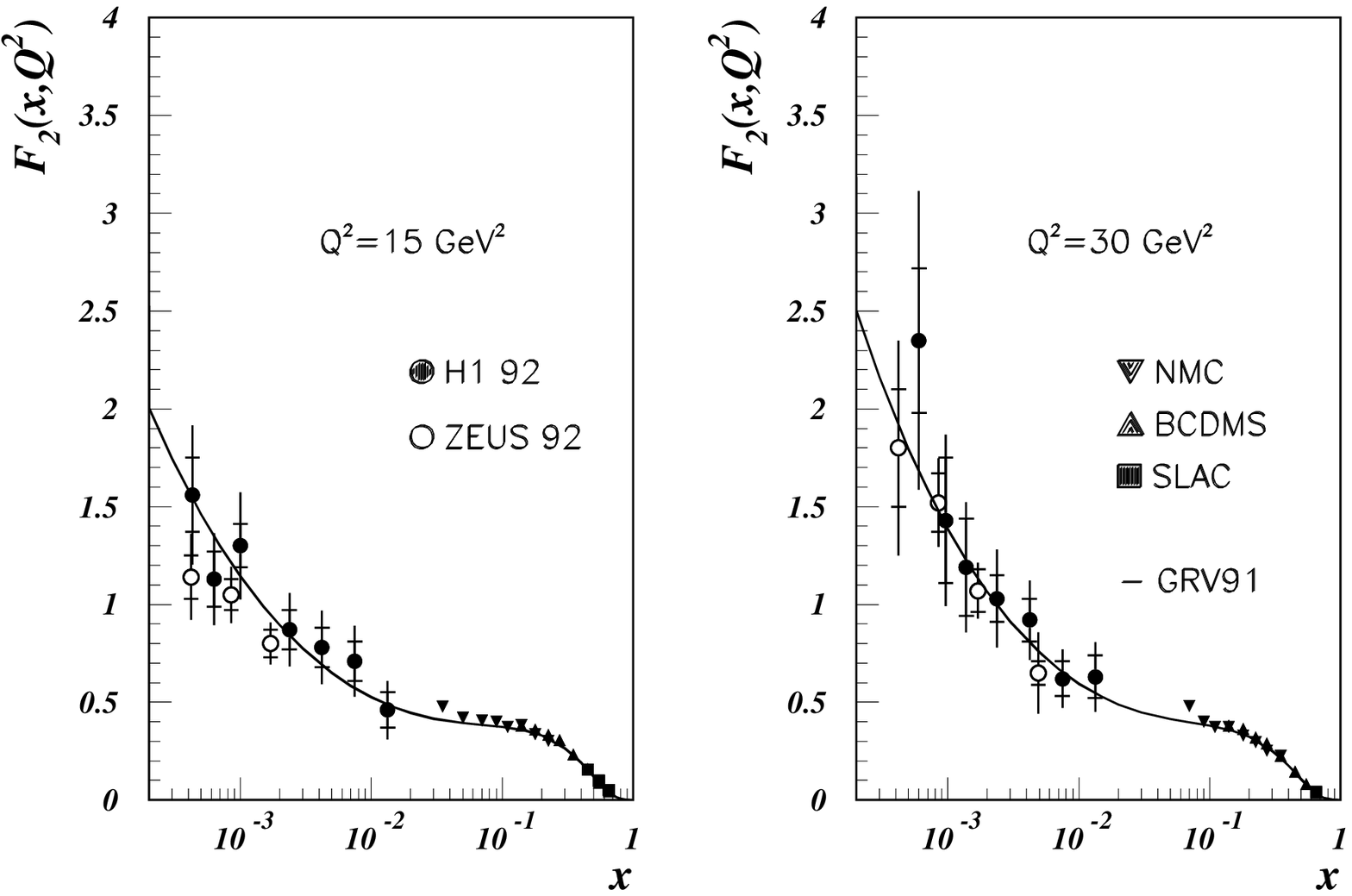,width=80mm,height=50mm}}
\caption{ } 
\label{F2}
\end{figure}

 In terms of the
partonic content of the nucleon the structure function, which reflets its
overall distribution, is given by 
\begin{eqnarray}  
F_2(x, Q^2)= \sum _f \, e_f^2 \, [\,q^f(x,Q^2)\,] \,,
\end{eqnarray}
 where the sum is over flavours weighted
by the respective squared charges ($e^2_f$). It is this function that is object
of main experimental studies, specially at HERA, at the small longitudinal
nucleon fraction of momentum, or small $x$ (see Fig. \ref{F2})
\cite{klein}.

The quark distribution function can be
shown to evolve as 
\begin{eqnarray} 
\frac{ \partial q^f(x, Q^2) }{ \partial \ln Q^2 } =
\frac{ \alpha_s }{ 2\pi } \int \frac{dx_1}{x_1} \, P_{qq}( \frac{x}{x_1})
\, q^f (x, Q^2) \,\,,
\end{eqnarray} where $P_{qq}=C_F \frac{1+z^2}{1-z}\left|_+\right.$ (with
self-energy corrections over the $k$ propagator) is one of the splitting
functions $P_{ij}$ ( $P_{gq}=C_F\frac{1+(1-z)^2}{z}$, etc), describing the
transition between the quark state $i$ to the quark state $j$, from fraction
of momentum $x_1$ to $x$. The above evolution refers to the non singlet quarks
distribution where sea quarks and gluon distribution are uncoupled
\begin{eqnarray} 
q_{NS}(x,Q^2) \equiv q_i(x, Q^2) - q_j(x, q^2)\,\,.
\label{qnsing} 
\end{eqnarray} 

 The singlet quark distribution is given by

\begin{eqnarray}
q_S(x, q^2) \equiv \sum_f \left[ \, q^f(x, Q^2) + \bar{q}^f (x, Q^2) \, \right]
\,\,,
\label{qsing}
\end{eqnarray}
where the gluon distribution is coupled
to the quark one.  Now the evolution equations, in the linear regime, read
for the quarks 

\begin{eqnarray} 
\frac{\partial q_S(x,Q^2)}{\partial \ln Q^2} =
\frac{\alpha_s(Q^2)}{2\pi} \left[ \int_x^1 \left( P_{qq} (\frac{x}{x_1})
q_s(x_1,Q^2) \right. \right. \nonumber \\  + \left. \left.
P_{qg}(\frac{x}{x_1})g(x,Q^2) \right) \right]  \,\,, 
\label{dglapq} 
\end{eqnarray}  

and for the gluon distribution
\begin{eqnarray} 
\frac{\partial g(x,Q^2)}{\partial \ln Q^2}= 
\frac{\alpha_s(Q^2)}{2\pi} \, \left[ \int_x^1 \left( P_{gq}(\frac{x}{x_1})
q_s(x_1,Q^2) \right. \right. \nonumber \\ 
+ \left. \left. P_{gg}(\frac{x}{x_1})g(x,Q^2) \right) \right]\,\,.
\label{dglapg} 
\end{eqnarray}

 The Eqs. (\ref{dglapq}) and (\ref{dglapg})
were independently derived by Dokshitzer, Gribov and Lipatov, and by Altarelli
and Parisi, known as DGLAP equations in leading order. The perturbative QCD
evolution, in the linear sector is governed by DGLAP equations, moreover a
suitable non perturbative inicial condition, extracted from the experiment for
a given boson virtuality.

It can be shown by
sucessive derivations that $q_{NS}(x, \varepsilon ) \sim \sum_n(\alpha_s
\varepsilon )^n$, $\varepsilon =\ln Q^2$, which corresponds to the emission
of $n$ gluons, showing that the DGLAP equations resum the leading $\ln Q^2$.
This can be understood as ladder diagrams with a strong ordering in transverse
momenta $k_{\perp}$, i.e., $Q_0^2 << k_1^2<< ...<<k_n^2 <<Q^2$. The scale
$Q_0^2$ is the cut, or transition value between perturbative and non
perturbative physics. It was shown by Gribov \cite{r6} that this result is
gauge independent once one considers the leading logarithm approximation.

At small-$x$ the gluons dominate, since $P^{(0)}_{gg}(z) \sim \frac{2N_c}{z}$,
and the parton distributions have  the general behavior $xp_i(x, Q^2) \sim
x^{-\lambda}$, $\lambda>0$. More likely for initial condition $xp_i(x,Q_0^2)
\sim Const$ and $ xp_i(x,Q^2) \sim exp \sqrt{\ln( \ln Q^2) \ln 1/x}$, known
as double leading logarithm approximation (DLA). From that is clear that
DGLAP predicts the increase of the gluon distribution function, and of the
structure function $F_2$ with the decreasing of $x$, whose relation in this
kinematical regime is given by  \begin{eqnarray} \frac{\partial F_2(x,Q^2)}{\partial
\ln Q^2}=\frac{\alpha_s(Q^2)}{2\pi} \sum_f e_f^2 xg(x, Q^2) \end{eqnarray}
being equal to $\frac{2\alpha_s}{9\pi}xg(x,Q^2)$ for $n_f=3$.  The DLA
implies strong ordering in $x$ and $k_T$, i. e.,
$x_1>>x_2>>....>>x_{i-1}>>x_i>>x$ and $k_{T_1}<<k_{T_2}<<...
k_{T_{i-1}}<<k_{T_i}<<Q^2$, the  resum of logs of $\alpha_s \ln(1/x)  \ln
Q^2$, having as region of validity $\alpha_s<<1$, $\alpha_s \ln(1/x) <<1$ and
$\alpha_s \ln (1/x) \ln Q^2 \approx 1$.

The resum of all leading logarithms
of Bjorken $x$, or the energy, is characteristic of the
Balitsky-Fadin-Kuraev-Lipatov (BFKL) equation. For very low $x$ values the
$\ln s$ becomes large and $\alpha_s \ln 1/x \approx 1$ and the DLA is not
valid. The BFKL evolution equation is proposed for an unintegrated gluon
distribution function  in the transverse momentum variable. Its solution grows
as a power of the center of mass energy $s$ with the consequent violation of
the unitarity bound \cite{r7} at very high energies. The cure for this problem
was not reached in the next to leading order calculation \cite{r8}, and is
still under research for instance, through the resumming of all BFKL Pomeron
exchanges \cite{r9}, for the cross section as well as for the structure
function.

The amplitude for the scattering quark-quark with one gluon
exchange in the $t$ channel at lowest order is given by ${\cal A}_0(s,t) \sim
s/t$, and for the next order in $\alpha_s$ the leading terms are given by $\ln
s$, resulting  ${\cal A}_1(s,t) \sim {\cal A}_0(s,t) \ln s$. Once one goes to
higher orders the number of contributing diagrams increases and the
calculation gains enormously in complexity \cite{r10}, and the usual procedure
is to introduce an effective vertex (Lipatov vertex). It results that the
general term is ${\cal A}_n(s,t) \sim {\cal A}_0(s,t) \epsilon^n(t) \ln ^n (s)
/ n$,  where $\epsilon(t)$ is a suitable function to take care of infrared
divergencies, docile under regularization, for instance, dimensional
regularization.

Clearly the BFKL evolution is summing the terms $\alpha_s^n
\ln ^n (s)$, where lower order logarithms are neglected. The result for the
amplitude is
\begin{eqnarray}
{\cal A}(s,t) = {\cal A}_0(s,t)
\sum_{n=0}^{\infty} \frac{\epsilon ^n(t) \ln ^n (s)}{ n } \approx
s^{\alpha (t)} \,\,, 
\end{eqnarray}
 with $\alpha (t) =1+\epsilon (t)$. In this
case, there is still the $\epsilon(t)$ infrared divergencies to be cured.

When just the singlet contribution in the $t$-channel is considered in the
BFKL formalism, meaning that only Pomeron exchange diagrams are taken into
account, the amplitude is given by 
\begin{eqnarray}
\frac{Im {\cal A}(s,t)}{s} & = &  \frac{{\cal G}}{2\pi^2} \int d^2 k_1 d^2 k_
2 \, \Phi_A(k_1,q) \nonumber \\  
& \times & \frac{F(y,k_1,k_2,q)}{k_2^2(k-q)^2}\, \Phi_B(k_2,q) \,\,,\end{eqnarray}
where ${\cal G}$ is the color factor for the process and $q$  is the
transfered momentum in the $t$-channel; the functions $\Phi_i$ are the impact
factors setting the coupling of $F$ to the external particles and finally the
function $F$ is  the perturbative gluon ladder. At leading order it consists
into the exchange of two gluons but the sum of all terms implies an integral
equation for $F$, that is infrared finite for a reggeized gluon ladder. This
behavior of the kernel of the BFKL equation is connected with the QCD Pomeron
and it is the resum of the leading logarithms $\ln s$.
 
 The solution of the
BFKL equation predicts the steep growth of the gluon distribution with
decreasing $x$ as well as the diffusion of the transverse momenta. As well as
DGLAP equations, the BFKL equation predicts the growing of the cross section
in the small-$x$ regime since the dynamics of this observable is related with
the gluon distribution function.
 
 From this very brief discussion on the
main issues of the linear formalisms for the dynamics of the parton
distributions, it gets clear the need of formal improving in order to
include the unitarity corrections  preserving the Froissart limit.
This important aspect of high energy physics was pointed out many years ago by
Gribov, Levin and Ryskin (GLR) in Ref. \cite{r11}. I will present in the
following the main attempts developed in the recent years towards a non linear
dynamics for high density QCD, as well as the high energy phenomenology
provided.

\section{The Question}
The main question that is addressed
once treating hdQCD is how to analytically separate small and large distance
contributions to high energy amplitudes in a properly gauge invariant
formalism. This corresponds to establish the hard and soft scales for the
process of interest and develop the physical meaningful method to introduce
the unitary corrections (UC) into the parton dynamics. 
 Once the
theoretical need for UC is established we should look for their signatures
analysing different observables.   Besides comparing the
predictions of the distinct formalisms it is required a common limit between
them, probably to be set by a saturation scale, $Q^2_S$. There exists mainly
two  non linear perturbative approaches \cite{Kovchegov,AGL,AGL1} and a
non perturbative one \cite{McLerran}. Although some progress has been made
towards their connection there is no common analytic solution for the gluon
distribution $g(x,Q^2)$ for  all kinematical range.

 The physics of hdQCD shakes the parton model and the
cherished concept of incoherence that is behind the standard calculations. It
seems that in order to control the increasing of the gluon function some
gluon recombination mechanism has to take place as the energy gets higher and
higher. 

This is a good point to remind that the analysis of the structure
functions has already given us  some surprises, and the previous
important one was the EMC effect \cite{r15}, that has as a main result
$F_2^A/AF_2^n \neq 1$. This difference is not predicted if one requires
complete incoherence of the partons, and reveals the presence of nuclear
effects in the structure functions where they were not expected. A large
literature is devoted to this phenomena, but it is interesting to point out
that the shadowing behavior noticed in $J/\Psi$ production could be nicely
described \cite{r16}, as well as the comparison with Drell-Yan processes
\cite{r17}, for the first data at small-$x$, considering the recombination
approach developed by Mueller and Qiu \cite{r18} which is based on the GLR
proposals.

Two main aspects are in order, the control of the increasing of the gluon
distribution function as an unitarity imperative and the appearance of nuclear
effects in high energy processes. If this aspect is relevant for fixed target
quarkonia production, it is strongly important for physics of HERA-A, RHIC and
LHC with nuclei.

\section{The High Density QCD Approaches}

The leading logarithm approximation DLA (related to DGLAP) and LL(1/x)
(related to BFKL) result both into linear evolution equations for the gluon
distribution function. The effect of summing large logs in high energy regime
implies the increase of the gluon distribution function $g(x,Q^2)$ as well as
the cross section once $x$ decreases. However, this result violates the
unitarity of the scattering matrix, a main theorem of relativistic Quantum
Field Theory, the Froissart theorem \cite{r7}, which states the cross section
cannot grow faster than $\ln ^2 s$. Translated to the DIS, this implies
increasing  restrictions to the structure function and/or total cross section,
say lower than $\ln^2 (1/x)$ and provides a limitation in the $x$ range to the
application of linear evolutions in order to get suitable results.

Intuitively we can associate $xg(x,Q^2)$ to the number of gluons into the
nucleon, $n_g$, per rapidity unity, $y=\ln (1/x)$, with transverse size of
order $1/Q$. In the hadron-nucleon interaction it is the virtual gluon that
probes the nucleon structure, in analogy with the eletroweak boson in DIS. The
virtual gluon-nucleon cross section is 
\begin{eqnarray}
\sigma_{G^*N}(x,Q^2)=\sigma_0\,xg(x,Q^2) \,\,,
\end{eqnarray} 
where $\sigma_0=\sigma_{G^*g\rightarrow X}=Cte\,\frac{\alpha_s(Q^2)}{Q^2}$, is
the total cross section of the virtual gluon, with virtuality $Q^2$, and
nucleon gluon interaction. Assuming $\sigma_0=\pi R^2_{HAD}$, then
$\sigma_0 xg(x,Q^2)$ corresponds to the area occupied by the gluons in a
nucleon. As $x\rightarrow 0$, this transverse area may be comparable, or even
bigger, than $\pi R^2_{HAD}$, following DGLAP or BFKL predictions for small
$x$ or small $Q^2$. Approaching this regime the gluons may begin to
superpose spatially in the transverse direction and to interact, behaving not
anymore as free partons. These interactions should slower, or even stop, the
intense growing of the cross section, fixing the limit $\pi R^2_{HAD}$ in
the small $x$ regime.

Introducing the function $\kappa$, with probabilistic interpretation

\begin{eqnarray}
\kappa = \sigma_0 \frac{xg(x,Q^2)}{\pi R^2}\,\,,
\end{eqnarray}
it is possible to estimate in which kinematical region one can expect
modifications in the usual evolution equations. So to say, for $\kappa << 1$
the system stays at $x$ and $Q^2$ where the usual evolution equations
(linear) are applicable, governed by individual partonic cascades, without
interactions among the cascades.

As $\kappa \approx \alpha_s$, partons from distinct cascades begin to interact
due to spatial superposition. This specific kinematical regime or the onset of
the recombination mechanism was first studied by Gribov, Levin and Ryskin
\cite{r11} almost twenty years ago, proposing the introduction of non linear
terms into the evolution equation.

\begin{figure}
\centerline{\psfig{file=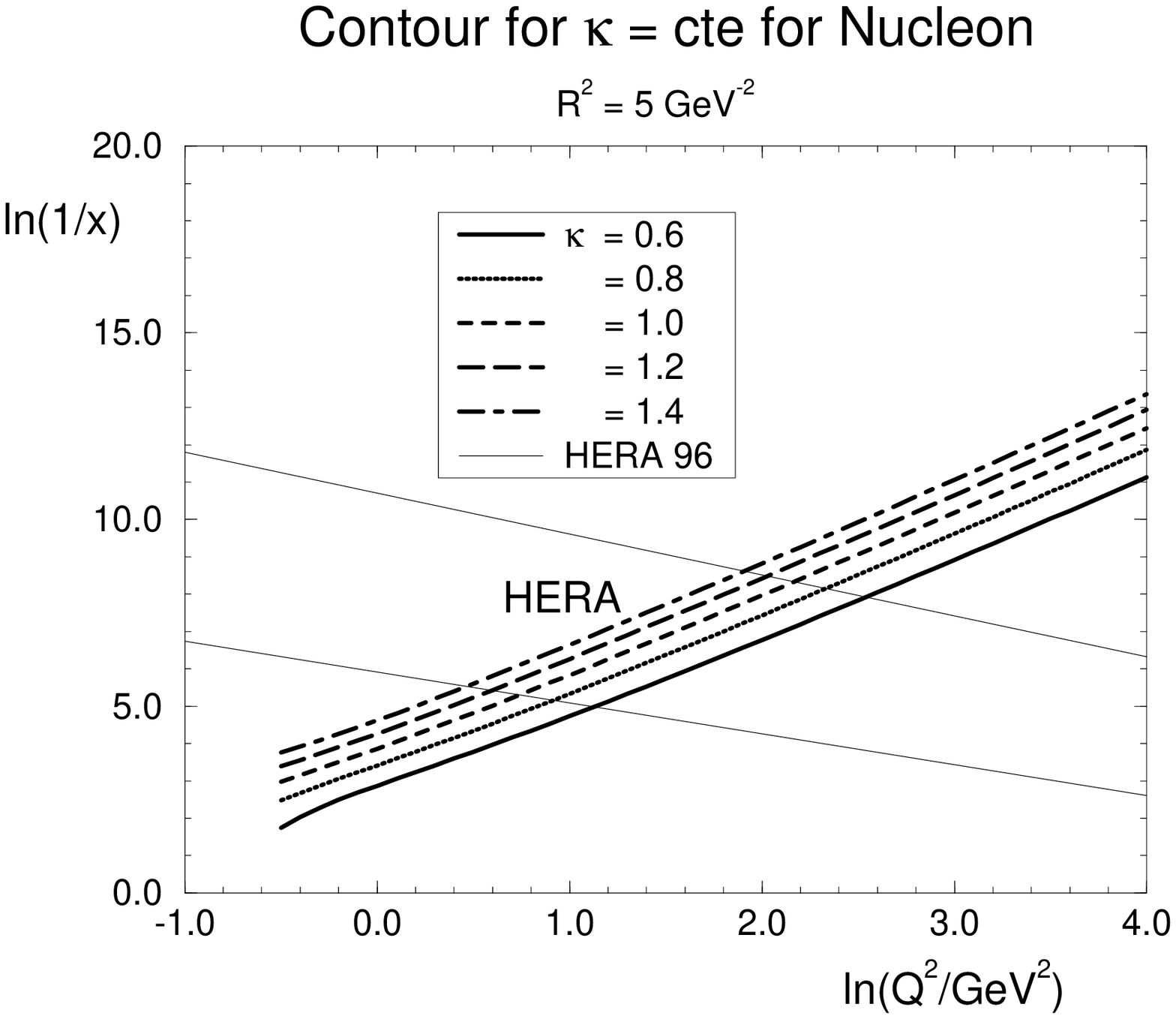,width=70mm}} 
\caption{ }
\label{kappa}
\end{figure}

The region of $\kappa \rightarrow 1$ was addressed more recently
\cite{AGL,AGL1}  (see Fig. \ref{kappa}) and experienced considerable
development on the theoretical side \cite{McLerran,Kovchegov}, also motivated
by HERA results and the great interest on RHIC and LHC future data. This is the
kinematical regime that requires the QCD dynamics for high partons density.
Although the coupling constant $\alpha _s$ is still small, allowing in
principle the use of perturbative methods, the system is so dense that
manifestation of non-linear effects are expected, and they are required to be
considered in a complete formalism.

The region of $\kappa \rightarrow 1$ corresponds
to partons in a non-equilibrium state and new methods are in order to treat
the collective phenomena. 

\vskip 1cm

\subsection{The GLR Formulation}

Gribov, Levin and Ryskin \cite{r11} introduced the mechanism of parton
recombination in perturbative QCD for high density systems, expressing this
as unitarity corrections included in a new evolution equation known as GLR
equation. In terms of diagrams it considers the dominant non-ladder
contributions, or multi-ladder graphs, also denoted fan diagrams.

The standard QCD evolution is represented by a cascade of partonic decays in
the nucleon. The photon interacts with a parton with fraction of momentum $x$
and virtuality $Q^2$, which is the last one of a chain where the partons get
slower and with bigger virtuality. The scale $Q_0$ sets the initial virtuality
and at the same time, the limit for perturbative QCD applicability, and $Q^2$
is the higher virtuality of the chain. In the transverse plane the partons
with low fraction of momentum stay in the lower part of the ladder and have
large transverse size; those with bigger virtuality are on the upper part of
the ladder and transversally smaller.

Following DGLAP, the number of partons of low fraction of momentum increases
very rapidly, which pictorically corresponds to bigger density of 
individuals in the same allowed area, in contrast with a more diluted system
at intermediate values of $x$, far away from the possibility of superposition.
The transition between these regimes should be characterized by a critical
value $x=x_{CRIT}$. The same can be argued through BFKL formalism, with the
difference that in this case the increasing of the partonic
distributions, takes place at a fixed transverse scale, although the evolution
presents the fluctuations in the transverse plane due to the characteristic
diffusion in BFKL.

It is important to emphasize that in both linear dynamics only the decay
processes are considered in the partonic evolution, however we expect that the
anihilation mechanism should contribute in the low $x$ regime,
providing some control of the increasing of the partons distribution function.
In the linear approach we consider one incident and two emergent partons to
construct the splitting functions for the decay processes. Now it is the case
to consider two incident partons and one emergent one, and to express the
recombination mechanism it is needed a formulation in
terms of the probability to recombine two incident partons. 

As a first approximation one considers the anihilation probability as
proportional to the square of the probability to find one incident parton,
introducing a non-linear behavior.

Taking $\rho = \frac{xg(x,Q^2)}{\pi R^2}$ as the gluon density in the
transverse plane, one has the general behavior: for splitting $1 \rightarrow
2$, the probability is proportional to $\alpha_s \, \rho$, for anihilation $2
\rightarrow 1$, the probability is proportional to $\alpha_s^2 \rho^2/Q^2$; 
where $1/Q^2$ stands for the size of the produced parton. For $x \rightarrow
0$, $\rho$ increases and the anihilation process becomes relevant. Considering
a cell of volume $\Delta \ln Q^2 \Delta \ln (1/x)$ in the
phase space allows one to write the modification of the partonic density as

\begin{eqnarray}
\frac{\partial^2\rho}{\partial \ln Q^2 \partial \ln 1/x} = \frac{\alpha_s
N_c}{\pi}\rho - \frac{\alpha_s^2 \gamma \pi}{Q^2}\rho^2\,\,,
\end{eqnarray}
where the coupling in the process is given by $\gamma$. Expressing in terms of
the gluon distribution the above equation is

\begin{eqnarray}
\frac{\partial^2 x g(x,Q^2)}{\partial \ln Q^2 \partial \ln 1/x }=
\frac{\alpha_s N_c}{\pi} xg- \frac{\alpha_s^2 \gamma}{Q^2 R^2}
[xg]^2 \,\,.
\end{eqnarray}

This equation is the GLR equation \cite{r11}. The already mentioned work of
Mueller and Qiu \cite{r18} gives $\gamma = 81/16$ for $N_c=3$.

\begin{figure}
\psfig{file=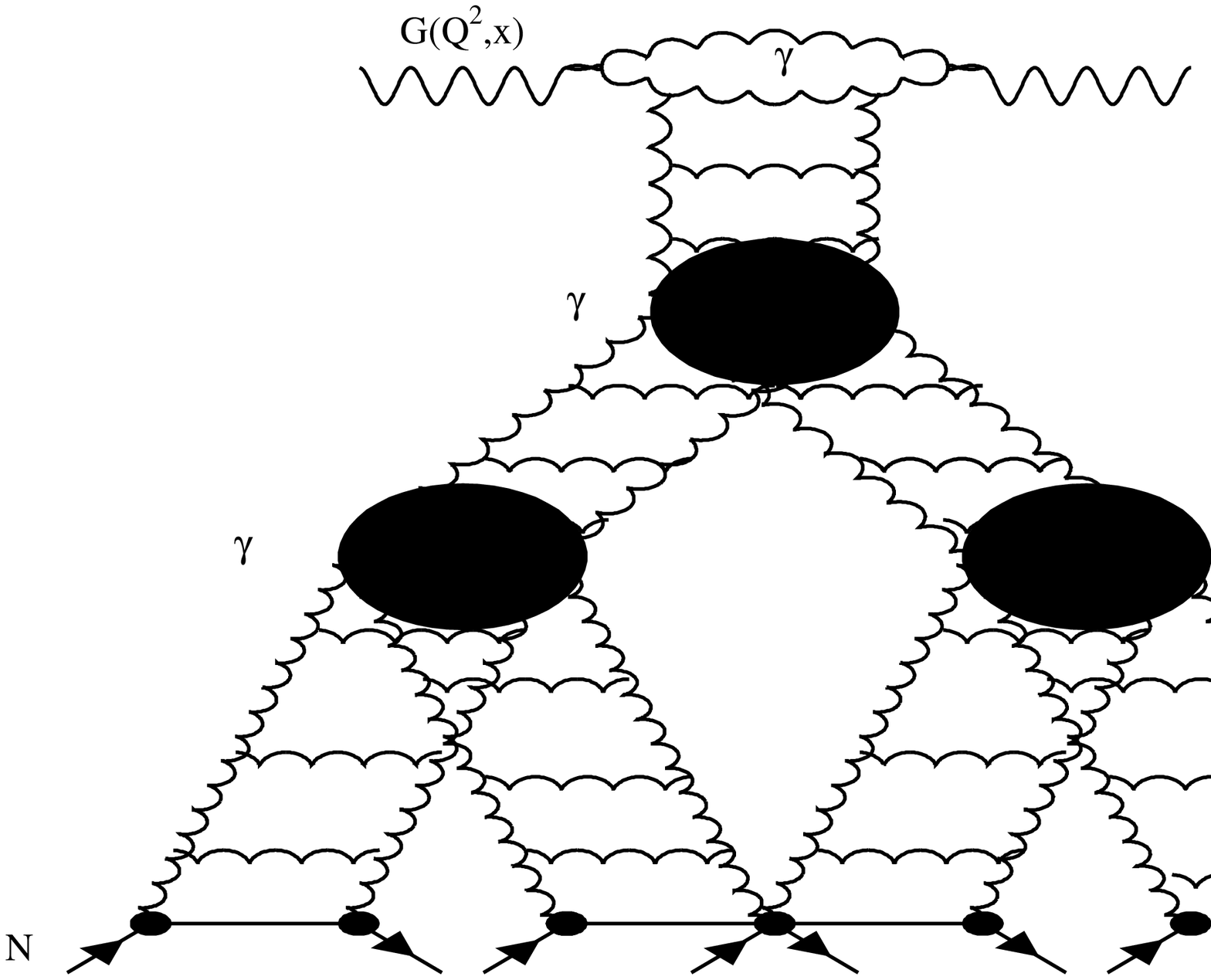,width=70mm}
\caption{ }
\label{fig44}
\end{figure}

The non-linear corrections correspond to a class of QCD Feynman diagrams,
called fan diagrams, formed by a gluon ladder with subsequent subdivisions in
gluon ladders, where the three ladders vertex is associated with the decay and
consists of a sum of several non planar diagrams. The overall result carries a
minus sign, which is important in order to control the growing of the parton
distribution once the fan diagrams become relevant, i.e., at low $x$. The
lowest part of the diagrams couples to the nucleon and the Eq. (14) resums all
class of the diagrams represented in Fig. \ref{fig44}.

As is clear from Eq. (14), the non-linear term reduces the growing of
$xg(x,Q^2)$ at low $x$, in comparison with the linear equations. It is also
predicted for the asymptotic region $x \rightarrow 0$ the saturation of the
gluon distribution, with a critical line between the perturbative
region and saturation region, setting its region of validity (meaning
independence of the gluon function with the energy). The subject of saturation
is very appealing and there are several attempts in the literature today with
distinct phenomenological approaches addressing this question \cite{r19}.

In the asymptotic limit one obtains $xg(x,Q^2)\left|^{GLR}_{SAT}
\right.=\frac{16}{27 \pi \alpha_s} Q^2 R^2$. Since the GLR only includes the
first non-linear term, although it predicts saturation in the asymptotic
regime its region of validity does not extend to very high density where 
higher order terms should contribute significantly.

\subsection{The AGL Formulation}

This approach developed by Ayala, Gay Ducati and Levin (AGL) \cite{AGL,AGL1},
intents to extend the perturbative treatment of QCD up to the onset of high
density partons regime, through the calculation of the gluon distribution
which is the solution of a non-linear equation that resums the multiple
exchange of gluon ladders, in double leading logarithm approximation (DLA).

\begin{figure}
\centerline{\psfig{file=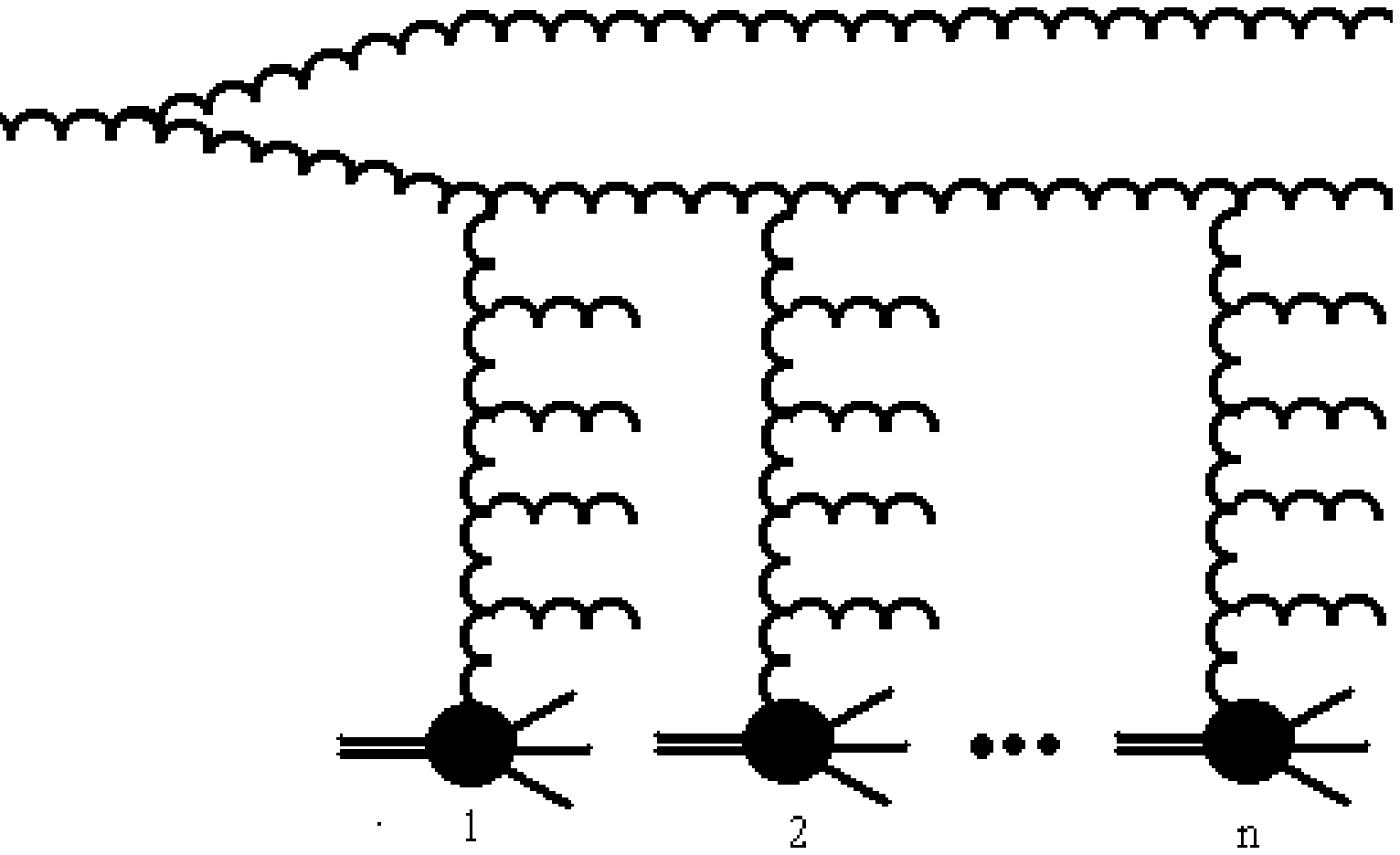,width=70mm}} 
\caption{ } 
\label{fig47}
\end{figure}

It is based on the development of the Glauber formalism for perturbative QCD
\cite{r20}, considering the interaction of the fastest partons of the ladders
with the target, nucleon or nucleus, since one of the main goals is to obtain
the nuclear gluon distribution $xg^A(x, Q^2)$. We considered a virtual probe
$G^*$ that interacts with the target in the rest frame, through multiple
rescatterings with the nucleons. In this reference frame the virtual probe
can be interpreted following the decomposition of the Fock states, and its
interaction with the target occours by the decay of the component $gg$, as
represented in Fig. (\ref{fig47}).

For small-$x$ this pair has a lifetime much bigger than the nucleus (nucleon) 
radius and the pair is separated by the fixed transverse separation $r_t$
during the interaction, which is represented by the exchange of a ladder of
gluons strongly ordered in transverse momentum.

The cross section for this process is given by
\begin{eqnarray}
\sigma^{G^* A}=  \int_0^1 dz \, \int \frac{d^2r_t}{\pi}\,
|\Psi_t^{G^*}(Q^2,r_t,x,z)|^2 \, \sigma^{gg \,+\, A} \,\,,
\end{eqnarray}
where $G^*$ is a colorless virtual probe with virtuality $Q^2,\, z$ is the
probe fraction of energy carried by the gluon and $\Psi_t^{G^*}$ is
the wave function of the transversely polarized gluon in the probe, $
\sigma^{gg\,+\,A}(z,r_t^2)$ is the cross section of the pair with the target, which was
proven for perturbative QCD by Mueller in Ref. \cite{r20,r21}

The lower limit estimation of UC is obtained through the
incoherent rescatterings of the gluon pair, with the constraint that only the
fastest partons of the ladders interact with the target. Introducing the
transverse impact parameter $b_t$ and a profile function for the nucleus
$S(b_t)$ we get
\begin{eqnarray}
\sigma^{G^* A}=  \int_0^1 dz \, \int_0^1 \frac{d^2r_t}{\pi}\,\int
\frac{d^2b_t}{\pi} |\Psi_t^{G^*}(Q^2,r_t,z)|^2 \nonumber \\ \,2 [1-
e^{\frac{1}{2} \sigma_N^{gg}(x^{\prime},4/r_t^2)S(b_t)} ] \,\,, 
\end{eqnarray} where $x^{\prime}=x/(zr_tQ^2)$, $S(b_t)$ may be taken as
$\frac{A}{\pi R_A^2}e^{-b_t/R_A^2}$ for a gaussian profile, $\sigma_N
^{gg}=\frac{C_A}{C_F}\sigma_N^{q\bar{q}}$, where $\sigma_N^{q\bar{q}}=
\frac{C_F}{C_A}(\frac{3}{4} \alpha_s(4/r_t^2))\pi ^2 r_t^2 x g (x, 4/r_t^2)$,
and $4/r_t^2$ is a cut for the nonperturbative region. For the virtual probe
with virtuality $Q^2$ the relation $\sigma^{G^* A}(x, Q^2)= (\frac{4\pi ^2
\alpha_s}{Q^2}) x g_A(x,Q^2)$ is valid.

In this approach the gluon pair emission is described in DLA of perturbative
QCD and from the Feynman diagrams of order $\alpha_s ^n$, it should be
extracted only the terms that contribute with a factor of order ($\alpha_s \ln
1/x \ln Q^2/Q^2_0)^n$. The interaction of the gluon pair with the target
operates through the exchange of a ladder which satisfies the DGLAP evolution
equation in the DLA limit.

It is a working hypothesis that in high energy the successive
rescatterings can be taken as independent allowing the employ of Glauber
formalism, in such a way using  the eikonal procedure for a relativistic
particle propagating in the target. 

Our master equation for the interaction of the $gg$ pair with the target is
known as the Glauber-Mueller formula and is 
\begin{eqnarray}
xg_A(x, Q^2) = \frac{4}{\pi^2} \int_x^1 \frac{dx^{\prime}}{x^{\prime}} \,
\int_{4/Q^2}^{\infty} \frac{d^2r_t}{\pi r_t^4} \, \int \frac{d^2 b_t}{\pi}
\nonumber \\  2[1- \sigma_N^{gg}(x^{\prime},4/r_t^2)S(b_t)] \,.
\end{eqnarray}
 
The pictorial representation of the space-time evolution of this formula is
given in Fig. (\ref{fig47}).


Once we perform the impact parameter integration using a gaussian profile
function we obtain
\begin{eqnarray}
xg_A(x, Q^2)= \frac{2 R_A^2}{\pi ^2} \, \int_x^1 \frac{dx^{\prime}}{x^{\prime}}
\, \int_{1/Q^2}^{1/Q^2_0} \frac{d^2r_t}{\pi r_t^4} \nonumber \\ 
\left[ C+
\ln(\kappa _G (x^{\prime},r_t^2)) + E_1(\kappa _G(x^{\prime},r_t^2)) \right]
\,\,, \end{eqnarray}
where $C$ is the Euler constant, $E_1$ is the exponential function and where
the $\kappa _G$ function was introduced as

\begin{eqnarray}
\kappa _G(x,r_t^2)=\frac{3 \alpha_s}{2 R_A^2} \pi r_t^2 xg(x,1/r_t^2) \,\,.
\end{eqnarray}

The expansion of Eq. (18) in terms of $\kappa _G$ gives as the Born term the
DGLAP equation in the small $x$ region, the higher order terms corresponding
to the unitarity corrections naturally implemented in this formalism.

\begin{figure}[t]
\centerline{\psfig{figure=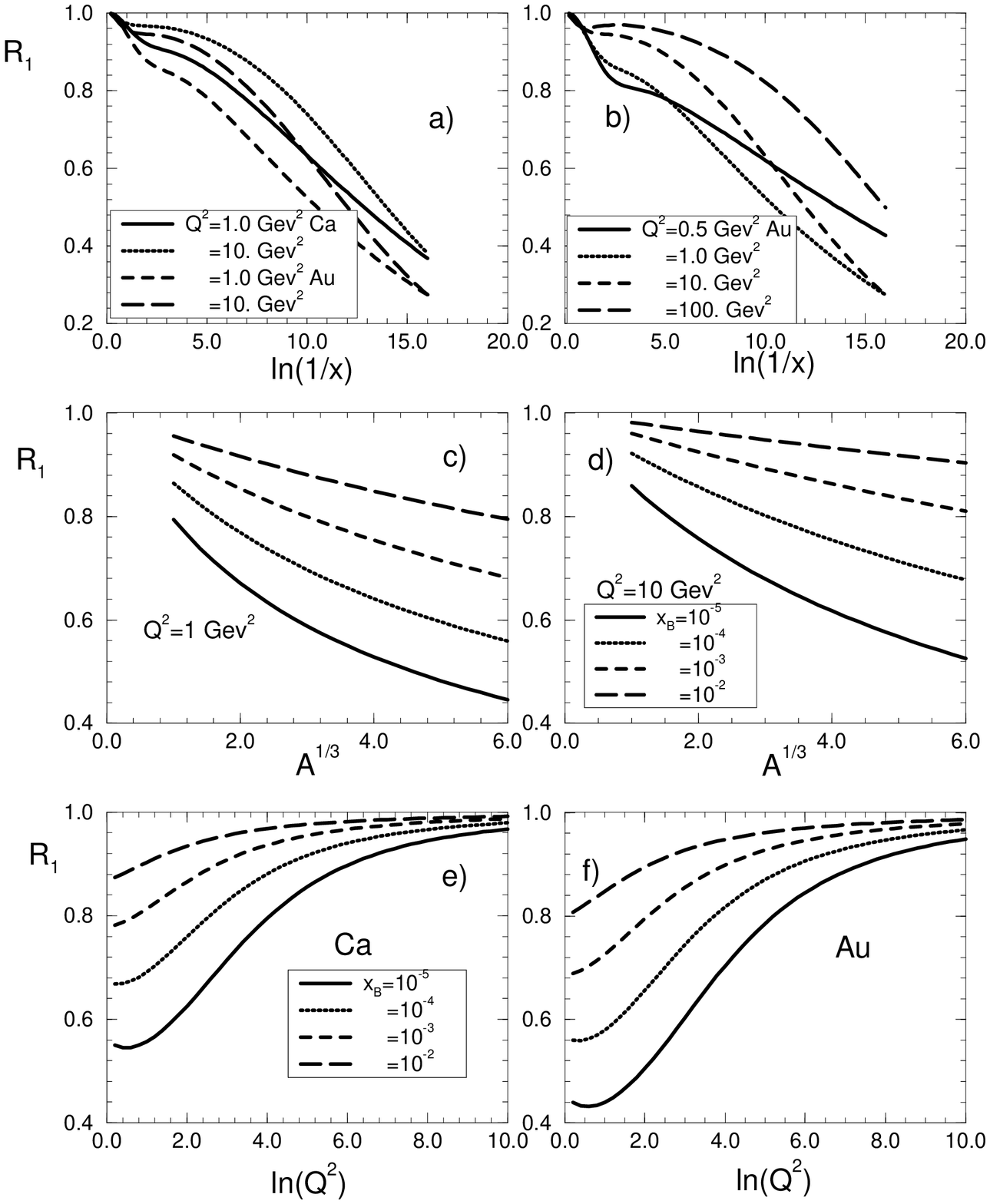,width=90mm,height=160mm}}
\caption{ }
\label{ayala}
\end{figure}

The estimation of the shadowing effect due to gluon recombination can be
immediately obtained studing the ratio $R_1=xg_A(x,Q^2)/Axg_N^{GRV}(x, Q^2)$
presented in Fig. (\ref{ayala}), where we calculate for $Ca$ and $Au$, and
analysed the behavior of this function in terms of $\ln (1/x)$, $A^{1/3}$ and
$\ln Q^2$. We used the GRV parametrization \cite{r23} and adapted the
calculation in order to have a larger domain of validity in $x$ using

\begin{eqnarray}
xg(x,Q^2)= xg_A[Eq. (18)] + A xg^{GRV}(x,Q^2) \nonumber \\
-A\, \frac{\alpha_s N_c}{\pi} \int_x^1  \int_{Q_0^2}^{Q^2} \frac{dx_1}{x_1}
\frac{dQ^{\prime \,2}}{Q^{\prime \,2}}x^{\prime}g(x^{\prime},Q^2) \,\,,
\end{eqnarray}
where $A\,xg^{GRV}(x, Q_0^2)$ is the initial condition. The same procedure
could be applied for another global parametrization based on DGLAP.

As expected the UC increase with $A$, and get smaller as $Q^2$ increases and
it is evident the importance of the effect as $x$ goes to small values. This
allows us to say that the UC should be included in the calculations related
with the nuclear gluon distribution function, and the obtained function $xg_A(x, Q^2)$
may be used to set the initial conditions for future experiments. For
instance in HERA-A, in processes $e^{\pm} A$ \cite{r24} the function $xg_A$
could be obtained indirectly and employed as an initial condition for the
hadronic high energy processes at RHIC and LHC.

The quarks and gluons distribution were also analysed for the nucleon in this
formulation \cite{AGL1}, as well as the structure function $F_2$ \cite{r25}.
The motivation for this generalization is the availability of HERA data. 

The free interpretation of the Froissart theorem for hadronic processes
requires a limit for the increasing of the cross section $\sigma_{\gamma ^*N}$
and $F_2$ with the energy so unitarity is not violated. Concentrating the
discussion on the $\kappa $ value, $\kappa = xg(x,Q^2) \sigma ^{gg}/(Q^2 \pi
R^2)=3 \pi \alpha_s xg(x,Q^2)/2Q^2 R^2$, which is the probability of gluons
interactions inside the partonic cascade, and $R$ is the radius of the nucleon
area occupied by the gluons, we were able to obtain $R^2= 5$ GeV$^{-2}$, and
that $\kappa$ reaches 1 at HERA, meaning the effects of shadowing should be
considered in the analysis \cite{AGL,AGL1,r26}. In the nucleon case, following
the same steps as before we obtain
\begin{eqnarray}
xg(x,Q^2)=\frac{4}{\pi^2} \int_x^1 \frac{dx^{\prime}}{x^{\prime}}
\int_{4/Q^2}^{\infty} \frac{d^2r_t}{\pi r_t^4} \int_0^{\infty} \frac{d^2
b_t}{\pi} \nonumber \\ 2\,[1-e^{-\frac{1}{2} \sigma_N^{gg} (x^{\prime},4/r_t^2)
S(b_t)}] \,\,, \end{eqnarray} 
and requiring the recovering of DGLAP at DLA we have
\begin{eqnarray}
\sigma_N^{gg} (x,4/r_t^2)=\frac{3\pi^2 \alpha_s}{4}r_t^2 x g(x,4/r_t^2) \,.
\end{eqnarray}

For the quarks, considering the scattering of a virtual photon that decays
into  a quark-antiquark pair, which interacts with the nucleon through the
exchange of a ladder we get 
\begin{eqnarray}
\sigma (\gamma^*)= \int_0^1 dz \int d^2 r_ t | \Psi (z, r_t)|^2 \,
\sigma_{tot}^{q \bar{q} + N}\,,\end{eqnarray}
where $\Psi$ is the wavefunction of the $q\bar{q}$ in the virtual photon
\cite{r21}. We obtain 
\begin{eqnarray}
\sigma_{tot}=\int d^2 b_t \,[1-e^{- \frac{1}{2} \Omega_{q\bar{q}}(x,r_t,b_t)}]\,.\end{eqnarray}

Here $\Omega _{q\bar{q}}$ is the opacity function that in the Glauber (or
eikonal) approach is equivalent to $\Omega = \frac{4 \pi^2
\alpha_s(Q^2)}{3Q^2} xg(x,Q^2)S(b_t)$. In doing so we are able to reproduce
DGLAP evolution for $\Omega <1$, and guarantee the validity of the formulation
also for the kinematical region where $\Omega >1$.

Taking $\Omega \rightarrow \infty$ and factorizing the $b_t$ dependence we
obtain the unitarity limit for the structure function, having for the $\ln
Q^2$ derivative of $F_2$ , $\partial F_2/ \partial \ln Q^2  < Q^2\,R^2/
3\pi^2$. Using GRV, the unitarity limit for HERA is reached for
$Q^2=Q^2_0=1-2$ GeV$^2$ ($y=\ln 1/x \sim 9$). Similarly, for gluons it is
$Q^2=1-2$ GeV$^2$ ($y=\ln 1/x \sim 7$) for HERA \cite{r26}.

With the aim to obtain a non-linear evolution equation containing the
unitarity corrections through the inclusion of all the interactions besides
the fastest parton from the ladder, we differentiate our master equation for
the gluon in $y=\ln 1/x$ and $\varepsilon=\ln Q^2$, obtaining
\begin{eqnarray}
\frac{\partial ^2 xg(y,\varepsilon )}{\partial y \partial \varepsilon }=
\frac{2\,Q^2 R^2}{\pi} \,[C + \ln (\kappa _G) + E_1(\kappa_G) ]\,,
\end{eqnarray}
where $\kappa _G^{DGLAP}(x,Q^2)= \frac{N_c \alpha_s \pi}{2Q^2 R^2} x
g^{DGLAP}(x,Q^2)$ for calculations.

In terms of $\kappa_G$ the main evolution equation is
\begin{eqnarray}
\frac{\partial^2 \kappa (y,\varepsilon )}{\partial y \partial \varepsilon } +
\frac{\partial \kappa (y,\varepsilon )}{\partial y}=  \frac{N_c
\alpha_s}{\pi} \,[C + \ln (\kappa _G) + E_1(\kappa _G) ]\,.
\end{eqnarray}
It should be mentioned that large distance effects are absorved in the initial
condition for the evolution, and situating in a conveninet region of $Q^2$
only short distance effects are present, meaning a perturbative calculation is
reliable.

Equations (25) and (26) were derived in Ref. \cite{AGL,AGL1}, and  refered
for simplicity as AGL equation. The main properties of this formulation are:
\begin{itemize}
\item all contributions from the diagrams of order $(\alpha_s y \varepsilon
)^2$ are resummed;
\item in the limit $\kappa \rightarrow 0$ the DGLAP evolution in DLA is fully
recovered;
\item for $\kappa < 1$, and not large, the GLR equation is recovered;
\item for $\alpha_s y \varepsilon \approx 1$ the equation is equivalent to the
Glauber formalism.   
\end{itemize}

\begin{figure}
\centerline{\psfig{file=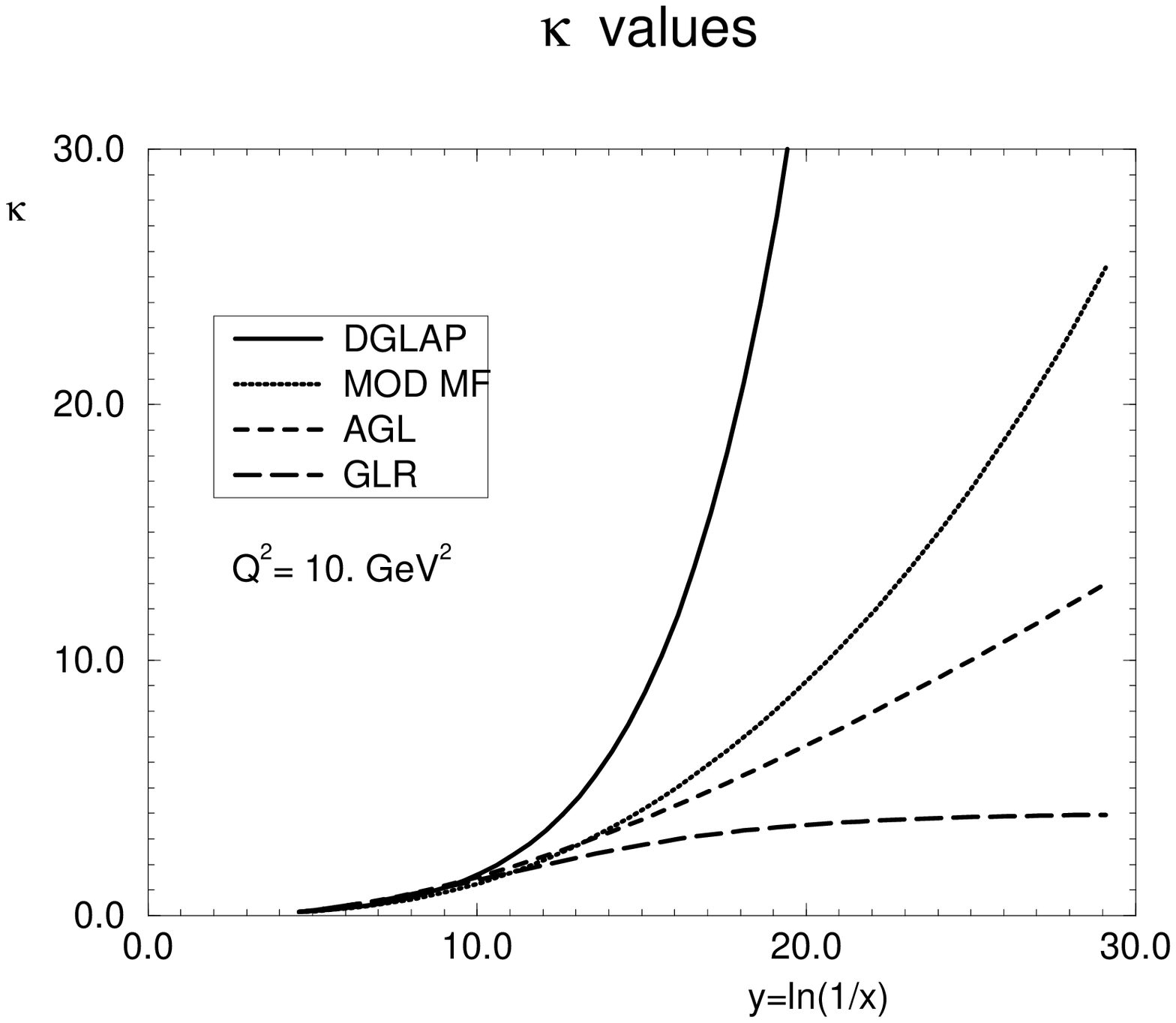,width=70mm,height=70mm}} 
\caption{ } 
\label{fig48}
\end{figure}

The UC are described for the different kinematical regions of $\kappa$ from
strictly perturbative QCD up to the onset of hdQCD. Non-perturbative effects
are not explicitly described and this is the object of a distinct formalism
$MV-JLKM$ \cite{McLerran} that we will briefly comment in a next subsection. In
Fig. (\ref{fig48}) the comparison between the solutions of the equations AGL,
GLR, DGLAP and Glauber-Mueller (MOD MF) formula is presented, where the control
of the growing of the gluon distribution once UC are considered is very
evident.

It was also obtained the asymptotic solution $\kappa > 1$ of the AGL equation
for fixed $\alpha_s$ \cite{AGL,AGL1} as well as for running $\alpha_s$
\cite{r27}. For high partonic density and $y>>y_0$ we obtain
\begin{eqnarray}
\kappa_G^{asymp} (y) = \frac{\alpha_s N_c}{\pi} \,y \ln y \approx
\frac{\alpha_s N_c}{\pi} y \,.
\end{eqnarray}

This solution is a good approximation for very small values of $x$ (${\cal O}
(10^{-8})$), related with THERA physics \cite{r24}, region of a very dense
parton system. In terms of the gluon function the asymptotic behavior is 
\begin{eqnarray}
xg(x,Q^2)= \frac{2\,N_c Q^2 R^2}{3\,\pi^2}\/ \ln(1/x) \,,
\end{eqnarray}
presenting a behavior softer than predicted by DGLAP, meaning a partial
saturation. 

For the running $\alpha_s$ the result is \cite{r27}
\begin{eqnarray}
xg(x,Q^2)= \frac{\varepsilon}{1+\varepsilon}\, \frac{2N_c Q^2 R^2}{3 \pi^2}
\ln (1/x) \,.
\end{eqnarray}
where $\varepsilon = \ln Q^2/ \Lambda _{QCD}^2$. The partial saturation is not
modified, and the main difference from the previous result occours for small
values of $\varepsilon$. This confirms the expectation that the UC are already
relevant before the corrections to leading order \cite{r26,muellerp}.

\subsection{The Kovchegov Formulation}

The unitarization problem in QCD was addressed as an extension of the dipoles
formalism for the BFKL equation by Kovchegov \cite{Kovchegov}.  This work
proposes a non linear generalization of BFKL equation, also addressed
previously in Ref. \cite{bal} by the use of OPE to QCD obtaining the
evolution of Wilson line operators. The scattering of a dipole (onium -
$q\bar{q}$) with the nucleon is described  by a cascade evolution
corresponding to the successive subdivision of dipoles from the father dipole.
Each dipole has multiple scatterings with the nucleons of the target, implying
multiple ladders exchange to be resummed in order to obtain the cross section
of the interaction of the dipole with the nucleus. As a result it is derived
the evolution equation having the unitarized BFKL Pomeron as solution, in the
LL($1/x$) approximation.

The scattering of the onium $q\bar{q}$ (dipole) with the nucleus in the rest
frame, takes place through a cascade of soft gluons, which once taken in the
$N_c \rightarrow \infty$ limit is simplified by the suppression of non-planar
diagrams. The gluons are replaced by $q\bar{q}$ pairs and the dipole Mueller's
technique for the perturbative cascade can be employed \cite{r30}.

\begin{figure}
\centerline{\psfig{file=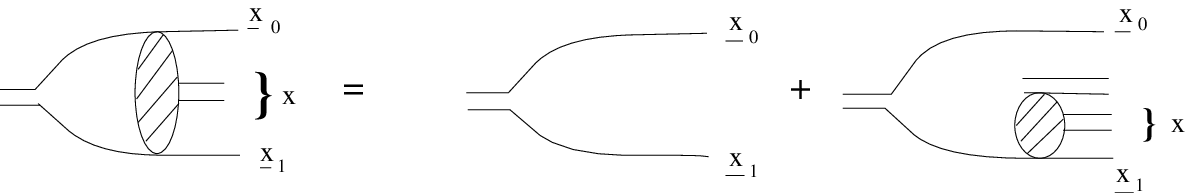,width=70mm,height=25mm}} 
\caption{ } 
\label{dipoles}
\end{figure}

The Kovchegov formulation, as the AGL, is a perturbative QCD calculation and
the  considered dipoles from the cascade interact independently with the
nucleus. The onium-onium frontal scattering has the cross section $\sigma = -
2\,{\cal I}m {\cal A}$, where the amplitude 
\begin{eqnarray}
{\cal A} = -i \int d^2x \int_0^1 dz \int d^2x_1 \int_0^1 dz_1\,
\Phi(\vec{x},z)\, F \, \Phi(\vec{x_1},z_1)\,, \nonumber
\end{eqnarray}
where $\Phi(\vec{x},z)$ is the square of the onium wave function, $\vec{x}$ is
the transverse separation of the $q\bar{q}$ pair, and $z$ is the longitudinal
fraction of momentum of the quark. For the exchange of only two gluons,
without gluon ladder evolution the function $F$ is \cite{r21}
\begin{eqnarray}
F^{(0)}(\vec{x},\vec{x_1})=- \frac{\pi \alpha^2_s (N_c^2-1)}{N_c^2} x^2_< \,(1+
\ln (\frac{x_>}{x_<})) \,,
\end{eqnarray}
where $x_>(x_<)$ is the biggest (smaller) between $|\vec{x}|$ and
$|\vec{x_1}|$. The two gluons approximation is energy independent, but for
high energy the contributions of order $(\alpha_s Y)^n$ should be included
($Y=\ln s/M^2$ is the rapidity and $M$ is the onium mass), since they generate
the perturbative cascade evolution. The dipole approximation introduces an
arbitrary number of soft gluons in the square of the onia wave function
$\Phi$, and keeping $F$ as an exchange of two gluons avoids to deal with the
reggeization of the gluons and the effective vertex. The transverse
coordinates of the quark and antiquark of an ultrarelativistic onium state in
$+$ direction are $\vec{x_0}=0$ and $\vec{x}$, and successively in the
evolution the next emitted gluon should be softer. We have $p-k$ and $k$, as
the momenta for the pair, and $z_1=\frac{k_1^+}{p^+}$ (in light-cone variables
\cite{r32}), having as wave function 
\begin{eqnarray}
\Psi^{(0)}(x_{01},z_1)=\int \frac{d^2 k_1}{(2\pi)^2} \, e^{i \vec{k_1}
\vec{x_{01}}} \Psi^{(0)}(k_1,z_1) \,,
\end{eqnarray}
where  $\vec{x_{01}}= \vec{x_1} - \vec{x_0}$, and $\Phi^{(0)}=|\Psi^{(0)}|^2$, keeping factorization
in this procedure. This allows to obtain the dipole
density, $n$, considering $x_{02}> \rho, \, x_{12}> \rho$, where $\rho$ is an
ultraviolet cut also implied by $C$ in the expression below
\begin{eqnarray}
&n&(x_{01},x,Y)= x \delta(x-x_{01})\exp \left[-\frac{2 \alpha_s
N_c}{\pi}\, Y\,\ln(\frac{x_{01}}{\rho}) \right]  \nonumber \\
& + & \frac{\alpha_s
N_c}{\pi^2} \int_C \frac{x_{01}^2 \, d^2x_2}{x_{02}^2 \, x_{12}^2} \, \int_0^Y
dy \exp \left[-\frac{2 \alpha_s
N_c}{\pi}(Y-y)\,\ln(\frac{x_{01}}{\rho}) \right] \nonumber \\
& \times & n(x_{12},x,y) \,,\end{eqnarray}
for $Y=\ln s/M^2$, which is represented in Fig. (\ref{dipoles}).

The next step is to obtain an evolution equation for the dipole density
assuming the propagation of the dipoles in the target is represented by the
function $\gamma_1(\vec{x}, \vec{b})$, where $b$ is the impact parameter, and
which should be added to the density $n_2$, equally convoluted with
$\gamma_2(\vec{x_1}, \vec{b_1},\vec{x_2}, \vec{b_2})$, etc.  Assuming no
correlation among the dipoles $\gamma_n(...)=\gamma_1(\vec{x_1},
\vec{b_1})...\gamma_1(\vec{x_n}, \vec{b_n})$, the cross section for the
interaction onium nucleus $N(\vec{x_{01}}, \vec{b_0}, Y)$ is then given by
\cite{Kovchegov}

\begin{eqnarray}
-N(\vec{x_{01}}, \vec{b_0}, Y) =\sum_i^{\infty} \int
n_i(x_{01},Y,\vec{b_1}, \vec{x_1},...,\vec{b_i},\vec{x_i}) \nonumber \\
 \times  \left[\gamma(\vec{x_1}, \vec{b_1})\frac{d^2x_1}{2\pi x_1^2} d^2b_1
\right]... \left[\gamma(\vec{x_i}, \vec{b_i})\frac{d^2x_i}{2\pi x_i^2} d^2b_i
\right] 
\end{eqnarray}

Finally, omitting some steps of the calculation \cite{Kovchegov} the evolution
equation for $N( \vec{x_{01}}, \vec{b_0},Y)$ is 
\begin{eqnarray}
& & N(\vec{x_{01}}, \vec{b_0},Y)  =  - \gamma(\vec{x_{01}}, \vec{b_0}) \exp
\left[-\frac{4\alpha_s C_F}{\pi} \ln(\frac{x_{01}}{\rho}) Y \right] + \nonumber
\\  & & \frac{\alpha_s C_F}{\pi^2} \int_0^Y dy \exp \left[-\frac{4\alpha_s
C_F}{\pi} \ln(\frac{x_{01}}{\rho}) (Y-y)  \right]\, \times \nonumber \\
& & \times \int_{\rho} d^2x_2 \frac{x_{01}^2}{x_{02}^2 x_{12}^2} \, \left[ \, 2
N( \vec{x_{02}}, \vec{b_0} + \frac{1}{2}\vec{x_{12}},y) \right. - \nonumber \\
& & \, - \, \left. N( \vec{x_{02}}, \vec{b_0}+ \frac{1}{2}\vec{x_{12}},y)N(
\vec{x_{12}}, \vec{b_0} - \frac{1}{2}\vec{x_{20}},y) \right] \,,  
\end{eqnarray} where $x_{ij}=x_i-x_j$, the size of the dipole whose quark has
transverse coordinate $x_i$, and the antiquark $x_j$, $\gamma (\vec{x_{01}},
\vec{b_0})$ is the propagator of the pair $q\bar{q}$ through the nucleus,
describing the multiple rescattering of the dipole with the nucleons within
the nucleus. We denote this equation as the $K$ equation. 

The physical representation is comparable with the approach
Glauber-Mueller since the incident photon generates a $q\bar{q}$ that
subsequently emits a gluon cascade further interacting with the nucleus. At
large $N_c$ limit the gluon can be represented by a $q\bar{q}$ pair, and  we
can expect in this limit and DLA that the gluon cascade could be interpreted
as a dipole cascade. Although beguinning the formulations with distinct degrees
of freedom both $K$ and AGL  resum the multiple rescatterings in
their respectives degrees of freedom, which allows to consider they should
coincide in a suitable common kinematical limit, which we will show later on.

In DLA, where the photon scale of momentum $Q^2$ is bigger than
$\Lambda_{QCD}^2$,  the $K$ equation simplifies to
\begin{eqnarray}
\frac{\partial N(\vec{x_{01}}, \vec{b_0},Y)}{\partial Y} = \frac{\alpha_s
C_F}{\pi} x_{01}^2 \, \int_{x_{01}^2}^{1/ \Lambda_{QCD}^2}
\frac{d^2x_{02}}{(x^2_{02})^2} \, \nonumber \\
\times  \left[ 2\,N(\vec{x_{02}}, \vec{b_0},Y)-
N(\vec{x_{02}}, \vec{b_0},Y)\,N(\vec{x_{02}}, \vec{b_0},Y)\right]\,,
\end{eqnarray} 
which is the evolution in transverse size of the dipoles from $x_{01}$ up to
$1/\Lambda_{QCD}$. Now deriving in $\ln (1/x_{01}^2\,\Lambda_{QCD}^2)$ results
\begin{eqnarray}
\frac{\partial ^2 N(\vec{x_{01}}, \vec{b_0},Y)}{\partial Y \partial \ln
(1/x_{01}^2\,\Lambda_{QCD}^2) } = \frac{\alpha_s C_F}{\pi}\,
\left[ 2 - N(\vec{x_{01}}, \vec{b_0},Y)\right] \nonumber \\
\times  N(\vec{x_{01}},
\vec{b_0},Y) \,.  
\end{eqnarray} 
setting that the successive emission of dipoles generates larger transverse
size for each higher generation.

The linear term reproduces BFKL at low density, and the quadratic term
introduces UC unitaryzing the BFKL Pomeron and the equation reproduces GLR
once we assume $N$ directly related to the gluon distribution function.

\subsection{The MV-JKWL Formulation}

In the MV-JKWL formulation \cite{McLerran,37} a very dense system is treated
in the light-cone and considering the light-cone gauge ($A^+ \equiv 0$), $x
\equiv q^+_{Gluon}/Q^+_{Nucleon}$. In Ref. \cite{McLerran} the gluons
distribution for small $x$ is proposed for a large nucleus where the degrees
of freedom are virtual quanta from a classical field generated by the color
charge of the valence partons (static sources). The approach is originally
non-perturbative and the nucleus is considered in the infinite momentum frame,
transfering the scale of the problem to $\Lambda = 1/\pi R^2 \, dN/dy$, where
$N$ is the density of gluons. For small $x$ and a large nucleus
$\alpha_s(\Lambda)$ is small allowing some perturbative calculation in this
effective lagrangian formulation for gluons condensates.

The density of gluons in momentum space is obtained in terms of the correlation
of the gluons fields, in the light cone gauge.  The intrinsic quantum
fluctuations are replaced by a classical average on the color charge ensemble.
The gluons distributions at a given virtuality $Q^2$ and $x$ is obtained from
the density of gluons in the momenta space $dN/ dq^+ q^2 \vec{q}$, which is a
function of the gluon condensate $<A_i^a(x^-,
\vec{x})\,A(x^{\prime},\vec{x^{\prime}})>$, being 
\begin{eqnarray}
xg(x,Q^2) \equiv \int^{Q^2} d^2 \vec{q} \, x \, \frac{dN}{dx q^2 \vec{q}} \,.
\end{eqnarray}

The gluons distributions, in this framework where a large number of color
charges generates a QCD vector potential, is obtained in lowest order by
solving the classical Yang-Mills equations, $D_{\mu}\,F^{\mu \nu}=j^{\nu}$.

Introducing a regulator in the valence partons current singularity by
considering the color density $\rho$ a function of rapidity, it was obtained
\cite{39} an analytical solution for the classical correlations, with the
property that for high transverse momentum the classical gluons distribution
obeys the Weizsacker-Williams form, and has its behavior softened as $\ln (k_
t^2/ \chi (y,k_t^2) )$. Here $ \chi = \int_y^{\infty} \mu^2(y,Q^2)$ is the
squared color charge per unity of area for rapidity bigger than $y$.

In the classical MV the non-linear effects are included in the charge density
$\rho$ solution of Y-M equations. The quantum corrections are to be
considered, and from Ref. \cite{40} the perturbative result for the gluons
distribution up to second order in $\alpha_s$ is given by
\begin{eqnarray}
\frac{1}{\pi R^2} \frac{dN}{dx d^2 k_ t} = \alpha_s \Gamma \frac{1}{xk_t^2}
\left[1+\frac{2\alpha_sN_c}{\pi} \ln (\frac{k_t}{\alpha_s \mu}) \ln
(\frac{1}{x}) \right] \,, \end{eqnarray}
where $\Gamma= \frac{\mu^2(N_c^2-1)}{\pi^2}$ and $\mu^2$ is the square of the
color charge average density (per unity of area). The additional effect of
including the hard gluon was treated in Ref. \cite{38}, resulting in the low
density limit the BFKL equation, and for high virtualities the DGLAP equation.
For high density a complete solution was not yet obtained. In Ref. \cite{41}
JKLW analysed their evolution equation in DLA obtaining a generalization of
GLR.

\begin{figure}
\centerline{\psfig{file=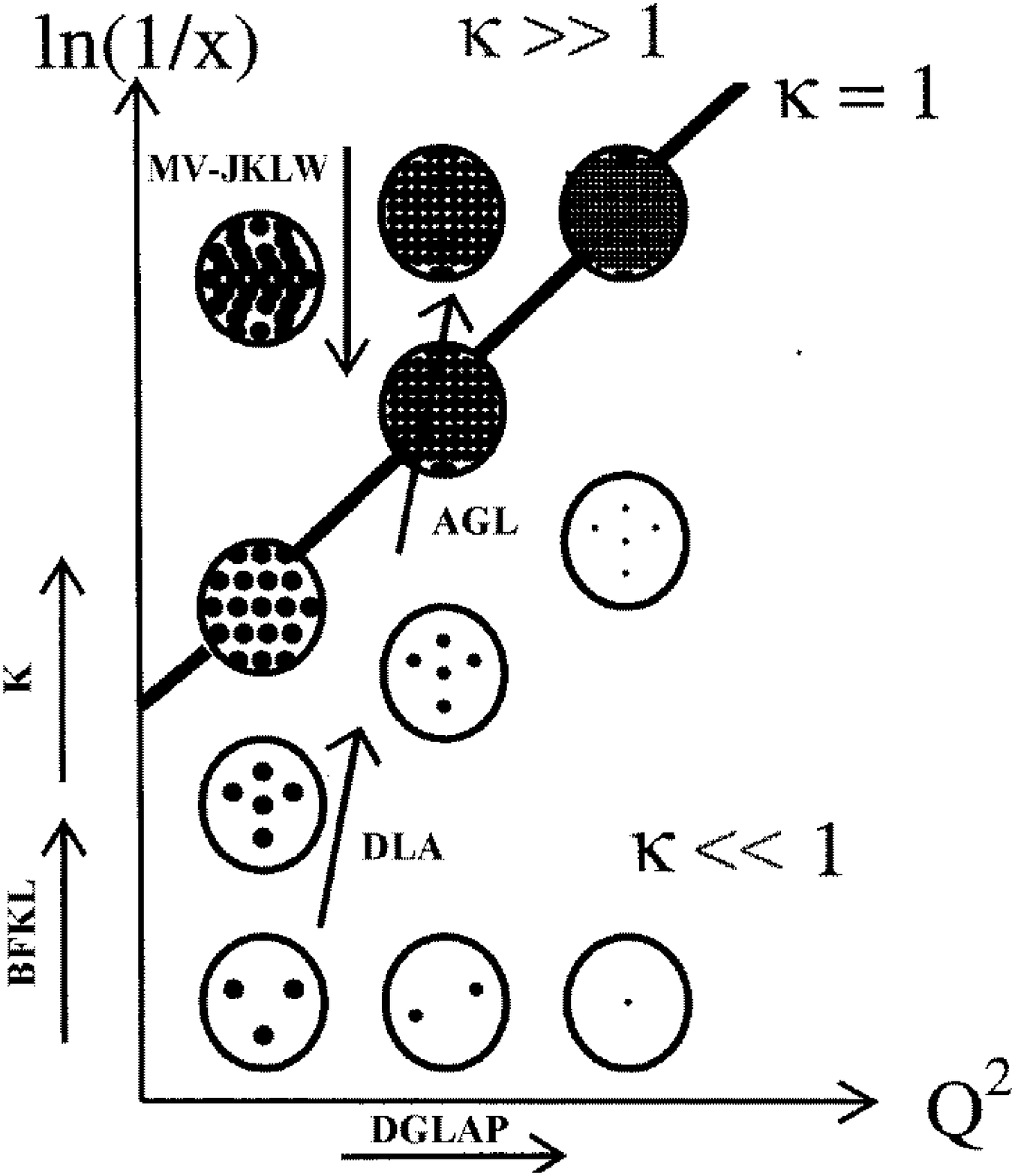,width=80mm,height=80mm}} 
\caption{  }
\label{fig8}
\end{figure}

As a summary of the formulations for hdQCD at present, the Fig. (\ref{fig8})
presents their different regions of applicability as fas as $\kappa$ in
concerned in the $\ln (1/x)$ versus $Q^2$ plane.

The main questions at this point can be:
\begin{itemize}
\item which is the most suitable form to introduce the UC ?
\item can we relate the distinct formulations in a common limit analytically ?
\item what do we look at the observables as a signature for the UC ?
\end{itemize}

The last two questions, I will briefly address in the rest of this
presentation following our personal contributions to this investigation. 

\section{Connection Among the Formulations}

The AGL equation was originally obtained from the Glauber-Mueller approach,
but it can be also derived from the dipole representation \cite{42}. We
obtained the cross section for the virtual probe $G^*$ with the nucleus
$\sigma^{G^*A}=\int_0^1 dz \int \frac{d^2r_t}{\pi} |\Psi_t^{G^*}|^2
\sigma^{gg+A}$, that can be expressed by means of the dipoles $q\bar{q}$ once
we remind $\sigma^{gg+A}=(C_A/C_F) \sigma^{q\bar{q} + A}$. Now in order to
estimate the UC the rescatterings of the  $q\bar{q}$ pair into the nucleus
should be considered, having in mind that \cite{r26}
$\sigma_N^{q\bar{q}}=\frac{C_F}{C_A} ( 3 \alpha_s(4/r_t^2)/4)\pi^2 r_t^2 x
g_N(x, 4/r_t^2)$, where $xg_N$ is the nucleon gluon distribution. The wave
function $\Psi^{G^*}$ calculated in \cite{r21,r26} is such that
$|\Psi_t^{G^*}|^2 = \frac{1}{z(1-z)}[(\epsilon^2 K_0(\epsilon t_t) - \epsilon
K_1(\epsilon r_t)/r_t )^2 + 1/r_t (\epsilon K_1(\epsilon r_t))^2 ]$, where
$\epsilon^2= Q^2 z(1-z)$, and the $K_i$ are the modified Bessel functions. For
small $z$ and $\epsilon r_t << 1$ we obtain
\begin{eqnarray}
xg_A(x,Q^2)= & & \frac{4}{\pi} \, \frac{C_A}{C_F}
\int_x^1\frac{dx^{\prime}}{x^{\prime}} \int_{4/Q^2}^{\infty} \frac{d^2r_t}{\pi
r_t^4} \nonumber \\
& \times & 2\, \left[\, 1-e^{-\frac{1}{2} \sigma_N^{q\bar{q}} S(b_t)}
\, \right]\,.\end{eqnarray}

$ $ From this equation we can obtain the AGL equation in the dipole
representation by differentiating in $y=\ln 1/x$ and $\ln Q^2/\Lambda_{QCD}^2$
having
\begin{eqnarray}
\frac{\partial^2 xg(x,Q^2)}{\partial y \, \, \partial \ln
Q^2/\Lambda_{QCD}^2} = C^{\prime} Q^2 \int
\frac{d^2 b_t}{\pi} \left[1-e^{-\frac{1}{2} \sigma_N^{q\bar{q}}S(b_t)} \right]
\,,\end{eqnarray}
valid in DLA, considering each gluon of the cascade as a $q\bar{q}$ in the
high $N_c$ limit, and where $C^{\prime}=2C_A/\pi^2 C_F$. For a central
collision and $S_{\perp}=\pi R^2$, and $S(0)=A/\pi R^2$
\begin{eqnarray}
\frac{\partial^2 xg(x,Q^2)}{\partial y \, \partial \ln
Q^2/\Lambda_{QCD}^2} = D Q^2 \left[1-e^{-\frac{2\alpha_s \pi^2}{N_c
S_{\perp}Q^2}xg_A} \right] \,,
\label{cil}
\end{eqnarray}
for $N_c=3$, $C_F=N_c/2$ at high $N_c$, and where $D= \frac{N_c C_F
S_{\perp}}{\pi^3}$.

The GLR for a cylindrical nucleus is immediately obtained from Eq. (\ref{cil})
by its expansion up to second order in $xg_A$. Again, for small UC only the
first term contributes which reproduces DGLAP in the DLA limit. Those results
are in Ref. \cite{42}

Now, the Eq. (34) is the K equation that in DLA, where the scale of momentum of
the photon $Q^2$ is higher than the scale of momentum of the nucleus
$\Lambda^2_{QCD}$, simplifies as 
\begin{eqnarray}
\frac{\partial N(\vec{x_{01}},\vec{b_0},Y)}{\partial Y} =
\frac{\alpha_s C_F}{\pi} x_{01}^2 \int_{x_{01}^2}^{1/\Lambda^2_{QCD}}
\frac{dx_{02}^2}{(x_{02}^2)^2} \nonumber \\
\times \left[ 2N(\vec{x_{02}},\vec{b_0},Y)-
N^2(\vec{x_{02}},\vec{b_0},Y)\right] \,.
\label{nn}
\end{eqnarray}

This equation considers the evolution of the dipoles from $x_{01}$ up to
$1/\Lambda_{QCD}$ in the transverse direction. Now deriving Eq. (\ref{nn}) in
$\ln (1/(x_{01}^2 \Lambda^2_{QCD}))$ we get
\begin{eqnarray}
\frac{\partial^2 N(\vec{x_{01}},\vec{b_0},Y)}{\partial Y \partial \ln
(1/x_{01}^2 \Lambda^2_{QCD}) } = \frac{\alpha_sC_F}{\pi}\left[\ 2-N \,\right] N
\,.\end{eqnarray}

We should relate now the function $N(\vec{x}_{01},\vec{b_0},Y)$ with the gluon
distribution function. For that we consider the structure function $F_2$ for
the nucleus as obtained in \cite{Kovchegov}, following \cite{r20}, and analized
in \cite{r25} for $b_t=0$, which is
\begin{eqnarray}
F_2^A(x,Q^2)=\frac{Q^2}{4\pi^2 \alpha_{em}}R^2 \int dz \int \frac{d^2
r_t}{\pi} |\Psi|^2 \nonumber \\
\times 2\left[1-e^{-\frac{\alpha_s C_F \pi^2}{N_c^2 S_{\perp}}r_t^2 Axg(x,
1/r_t^2)} \right]\,.  
\end{eqnarray}

This estimates the UC for the nuclear structure function, for central
collisions in the DLA limit in the Glauber-Mueller approach.

Considering the unitarity corrections due to the multiple rescattering of the
$q\bar{q}$ pairs with the distinct nucleons into the nucleus, from the just
obtained expression for $F_2^A$, it results the relation 
\begin{eqnarray}
N(\vec{x_{01}},\vec{b_0}=0,Y)= 2\left[1-e^{\frac{-2\alpha_s C_F \pi^2}{N_c^2
S_{\perp}} x_{01}^2 A xg(x,1/x_{01}^2)} \right]\,,
\end{eqnarray}
where $x_{01}=x_0-x_1=r_t$, $Y=\ln (s/Q^2)=\ln (1/x)$ establishing a
connection between the cross-section of the $q\bar{q}$ pair and the gluon
structure function in DLA limit.

We are in good terms to verify the connection among the K and AGL formulations
since we already obtained the AGL equation in the dipole formulation, Eq.
(40),  the K equation in the DLA limit, Eq. (42), the cross section of the
pair through the dipole density from K and the nuclear gluon distribution
function, Eq. (45).

Having Eq. (45) in Eq. (41) and for $x_{01} \approx 2/Q$, as in
\cite{McLerran}, we obtain
\begin{eqnarray}
\frac{ \partial^2 xg_A(x,Q^2)}{\partial y \partial \ln
(Q^2/\Lambda_{QCD}^2)} = D \,Q^2
\left[1-e^{-\frac{2 \alpha_s \pi^2}{N_c s_{\perp} Q^2}xg_a} \right]
\,, \end{eqnarray}
result already obtained, and that gives GLR as a limit. 

Our comparison has physical meaning for dipoles with small transverse sizes
and for the above connection among $N$ and $xg_A$ [Eq. (45)].

In Refs. \cite{38,39} it was applied the Wilson renormalization group to the
model of McLerran and Venugopalan. The non-linear evolution equation then
obtained deals with the weight function of the color charge densities, valid
at leading order  $\alpha_s$ and for densities up to $1/\alpha_s$. The
complete analytical solution is not yet obtained but some limits are
discussed. At low densities BFKL is recovered, and then at DLA at large $Q^2$
DGLAP is recovered. In the work \cite{41} is  proposed  the equation
\begin{eqnarray}
\frac{\partial^2 xg(x,Q^2,b_t)}{\partial y \partial
\varepsilon}= \beta \, Q^2 \left[1-\frac{1}{x}exp(1/\kappa)
E_1(1/\kappa) \right]\,, 
\end{eqnarray}
where $\beta= N_c(N_c-1)/2$ and $\kappa
(x,Q^2,b_t)=2\alpha_s/ \pi (N_c-1)Q^2 xg(x,Q^2,b_t)$.

For large $\kappa$, a factorized $b_t$ dependence and considering a central
collision we obtain for this equation
\begin{eqnarray}
\frac{\partial^2 xg(x,Q^2)}{\partial y \partial
\varepsilon}= \beta  R^2 Q^2 \,,
\end{eqnarray}
which solution is 
\begin{eqnarray}
xg(x,Q^2)= \beta  \pi R^2 Q^2 \ln(1/x) \,,
\end{eqnarray}
presenting the same $Q^2$ and $x$ behavior as the asymptotic solution for AGL.
The main point is the partial saturation of the gluon distribution presented
in both formulations in the asymptotic region. A connection among those two
formulations in a more broad kinematical region is still an open question. 

The asymptotic behavior of the structure function also required our attention.
Considering the relation of $\sigma^{q\bar{q}}$ and $xg(x,Q^2)$ we can write
\cite{r25}

\begin{eqnarray}
F_2(x,Q^2)=\frac{2\alpha_s}{9\pi} \int_{Q_0^2}^{Q^2} \frac{dQ^2}{Q^2}
xg(x,Q^2)
\end{eqnarray}
which is a leading twist equation, with limited application for high densities,
due to higher twist terms related with the UC.

Using the solution of AGL in the asymptotic regime as input in the above
equation we obtain $F_2(x,Q^2)\simeq \frac{\alpha_s}{\pi^3} R^2 Q^2 \ln
(1/x)$, which
again presents partial saturation, meaning the Froissart limit is not violated
\cite{46}. Analogous result was obtained by Kovchegov \cite{47} employing the
solution of the K equation \cite{Kovchegov}. We obtained that the asymptotic
behavior of $F_2$ is a general characteristic that appears to be independent
from the approach that is used \cite{r27}.

Assuming the asymptotic behavior of the gluon function is
$xg(x,Q^2)=2Q^2R^2/3\pi \alpha_s$, it implies saturation for $F_2$ ($\sim R^2
Q^2$)  for very small $x$. However this result should be taken with caution
since it is valid in a kinematical region where higher order in the partonic
density are not significative. The subject of saturation is a tricky one and
it seems  we are far from establishing its features in a solid
theoretical basis \cite{r19}. Important contributions to these challenging
aspects of hdQCD are to be found in Mueller \cite{r19} for the theoretical
discussion and Golec-Biernat and W\"usthoff \cite{r19} for a phenomenological
application.

In \cite{45,46} we were able to show that 
\begin{eqnarray}
F_2(x,Q^2)=\frac{R^2}{2\pi^2} \sum_i e_i^2 \int_{1/Q^2}^{1/Q_0^2} \frac{d^2
r_t}{\pi r_ t^4} \left[C + \ln \kappa_q + E_1(\kappa_q) \right]\,, \nonumber \\
\end{eqnarray}
where $\kappa_q=4/9 \kappa_g$. From that we can estimate the UC for $F_2$ in
the DLA limit. For large $\kappa_q$, and using the asymptotic solution of AGL,
we obtained  \cite{r27}
\begin{eqnarray}
F_2(x,Q^2) \simeq \frac{R^2 Q^2}{3 \pi^2} \ln [\frac{4\alpha_s}{3} \ln (1/x)]
\,,\end{eqnarray}
when higher twist terms are considered in $F_2$. This is a softer behavior,
but in both cases there is no violation of the Froissart limit. This above
equation was not studied in K or MV-JKLW approaches.

$ $ From the already obtained results it follows the identity \cite{nova}
\begin{eqnarray}
\frac{\partial F_2(x,Q^2)}{\partial \ln Q^2}=F_2(x,Q^2)\,,
\end{eqnarray}
as an important signature of the asymptotic regime of QCD for dense systems.
It is relevant to mention that for the same center of mass energy this regime
is reached for nucleus for smaller partonic densities than in the nucleon
case, since $\kappa_A=A^{1/3} \kappa_N$. 

\section{Phenomenology}

$ $ From the Glauber-Mueller formalism for high dense partonic systems was
demonstrated the AGL equation and its asymptotic behavior. It was also
obtained the nucleon and the nuclear gluon distribution function as well as
the respective structure functions and derivatives. This formulation
incorporates the UC required by the Froissart bound, through a non linear
dynamics.

In this section the behavior of the main observables obtained in $ep$
collisions, and  relevant for $eA$ collisions, will be analysed with the goal
to shed some light in the subject of UC. 

For $ep$ we studied the behavior of the proton structure function $F_2$, its
derivative $\frac{\partial F_2}{\partial \ln Q^2}$, the charmed component of
the structure function $F_2^c$, and the longitudinal distribution function,
$F_L$ \cite{45}. 

There is a large amount of data from HERA to motivate a detailed study of
these observables directly connected with the gluon distribution function. As
previously demonstrated the gluon distribution is modified in a unitarity
corrected formulation, meaning those observables should be affected.

We were lucky to show that the  $\frac{\partial F_2}{\partial \ln Q^2}$, the
$F_2^c$ are clearly modified. Also, the $eA$ analysis provides stricking
results for the nuclear structure function $F_2^A$ and its derivative, as an
important signature of the UC corrections. These results are important since
they are a  prediction both for HERA-A and for $e$-RHIC, in which a high dense
parton system should be formed.

The increasing of $F_2$ in HERA in the small $x$ region
($10^{-2}>x>10^{-5}$) is observed even for small virtualities ($Q^2 \approx 1$
GeV$^2$). Taking $F_2 \sim x^{-\lambda}$, for small $x$ data is compatible
with $\lambda = 0.15$ ($Q^2=0.85$ GeV$^2$) up to $\lambda =0.4$ ($Q^2=20$
GeV$^2$).  This is described by DGLAP  with suitable input initial condition
for $Q^2$ and distributions by different groups \cite{r23,55}. It will conduct
to the idea UC are not observable in the HERA kinematical range. We have shown
that the structure function is too inclusive in the gluon function to clearly
explicitate the UC.

We arrived at a different conclusion applying AGL to $\frac{\partial
F_2}{\partial \ln Q^2}$, $F_L$ and $F_2^c$, all observables directly
associated with the gluon function.

\begin{figure}
\centerline{\psfig{file=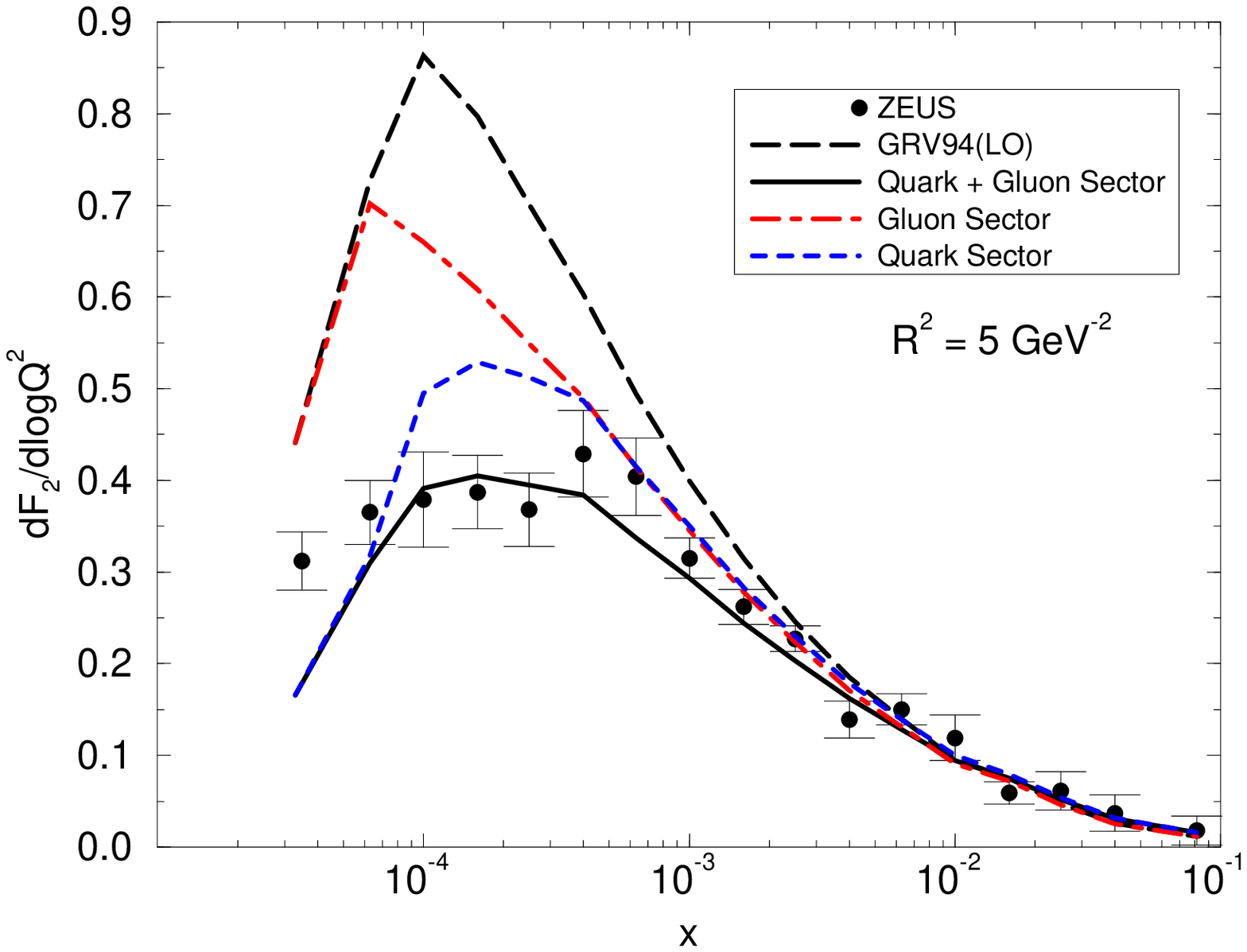,width=70mm}} 
\caption{  }
\label{derivada}
\end{figure}

The derivative of $F_2$ is

\begin{eqnarray}
\frac{\partial F_2}{\partial \ln Q^2}=\frac{R^2 Q^2}{2 \pi^2} \sum_i
e^2_i[\,C+ \ln (\kappa _q) + E_1 (\kappa_q) \,]\,,
\end{eqnarray}
that we solved using the same procedure as Eq. (20). The usual
parametrizations \cite{r23,55} do not include the UC for the gluons
explicitly. We use Eq. (20) for $A=1$ as input, and we obtain the corrections
from both sectors quark and gluons. The last one gains in importance as $Q^2$
increases. In Fig. (\ref{derivada}) the results for  $\frac{\partial
F_2}{\partial \ln Q^2}$ are presented for $R^2=5$ GeV$^2$. For the complete
discussion we refer to \cite{53}. The UC for  both sectors are able to describe
properly the data including the turn-over. Our conclusions is this is a good
observable to evidentiate the presence of the UC. This question was addressed
also in \cite{56} calculating the suppression factors separately.

We believe that the UC should be extracted from data related to observables
that are directly dependent of the gluon function. The longitudinal structure
function $F_L$ is a difficult measurement requiring distinct values of the
center of mass energy, meaning different energy beams. An alternative is
to consider the radiation of a hard photon from the incident electron,
reducing the center of mass energy. If this should be done we can study
$F_L(x,Q^2)$ in the small $x$ region \cite{45}.

\begin{figure}
\centerline{\psfig{file=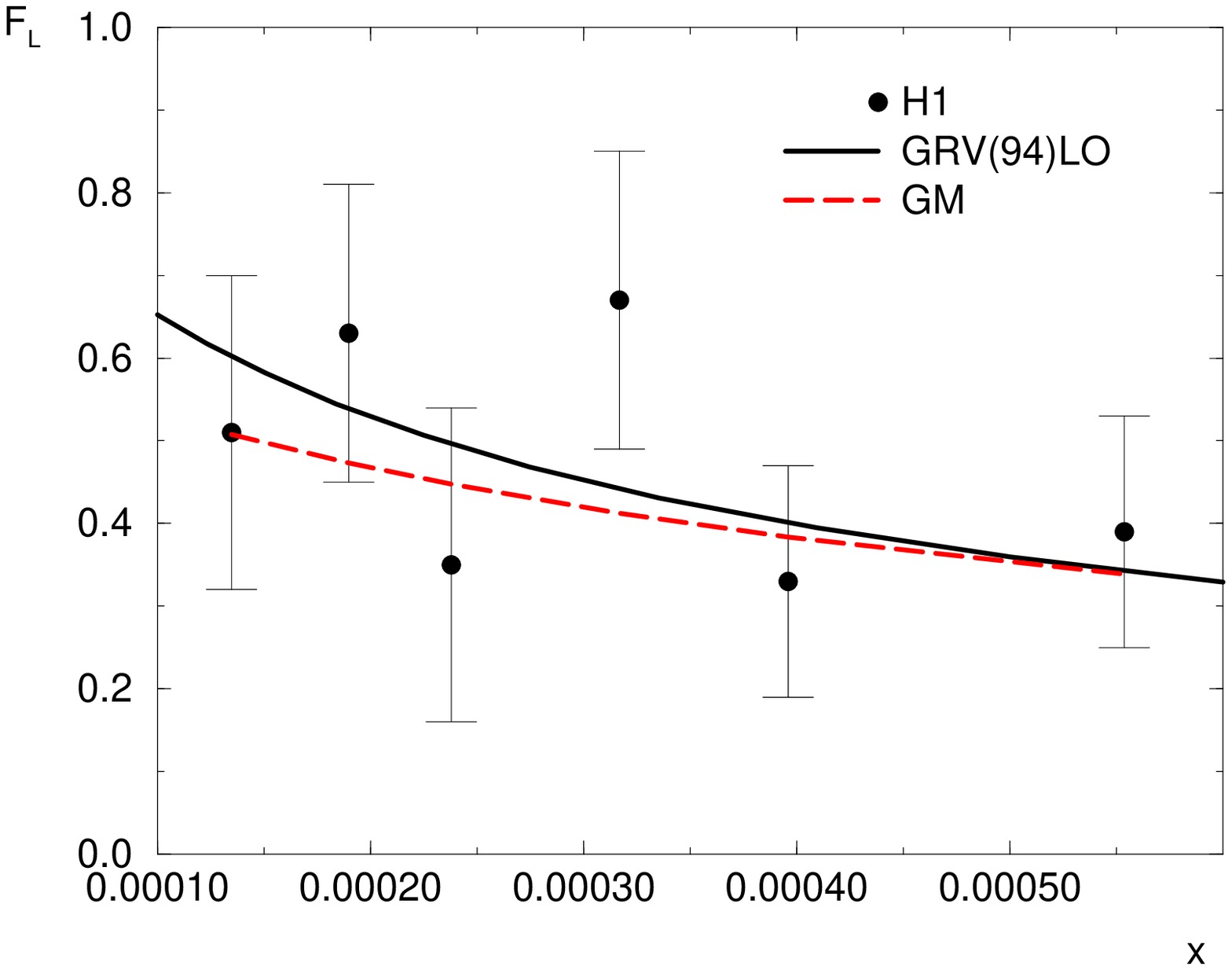,width=70mm}} 
\caption{  }
\label{fig76}
\end{figure}

Expressed considering the quarks transverse momenta due to gluon radiation,
the longitudinal structure function reads
\begin{eqnarray}
F_L(x,Q^2)=\frac{\alpha_s(Q^2)}{2\pi} x^2 \, \int_x^1 \frac{dy}{y^3} \left[
\frac{8}{3} F_2(y,Q^2) \right. \nonumber \\
\left. + \, 4 \, \sum_f e_f^2 (1-\frac{x}{y})yg(y,Q^2)
\right]\,, \end{eqnarray}
where $y=Q^2/sx$ and the dependence on the gluon distribution is explicit,
meaning this function should be sensitive to unitarity corrections in HERA
kinematical region. Our results for small $x$ region are in Fig. (\ref{fig76})
\cite{45} compared with the H1 data \cite{57}. Although it seems to be a good
observable to evidentiate the   UC the available data do not allow any definite
conclusion for now.

A problably more promissing observable is the rate
$R_F=F_2^c(x,Q^2)/F_2(x,Q^2)$, where $F_2^c$ is the charmed component of the
structure function. Considering the approach of boson-gluon in order to create
the $c\bar{c}$ pair we obtained in the Glauber-Mueller formalism the ratio
$R_F$. 

\begin{figure}
\centerline{\psfig{file=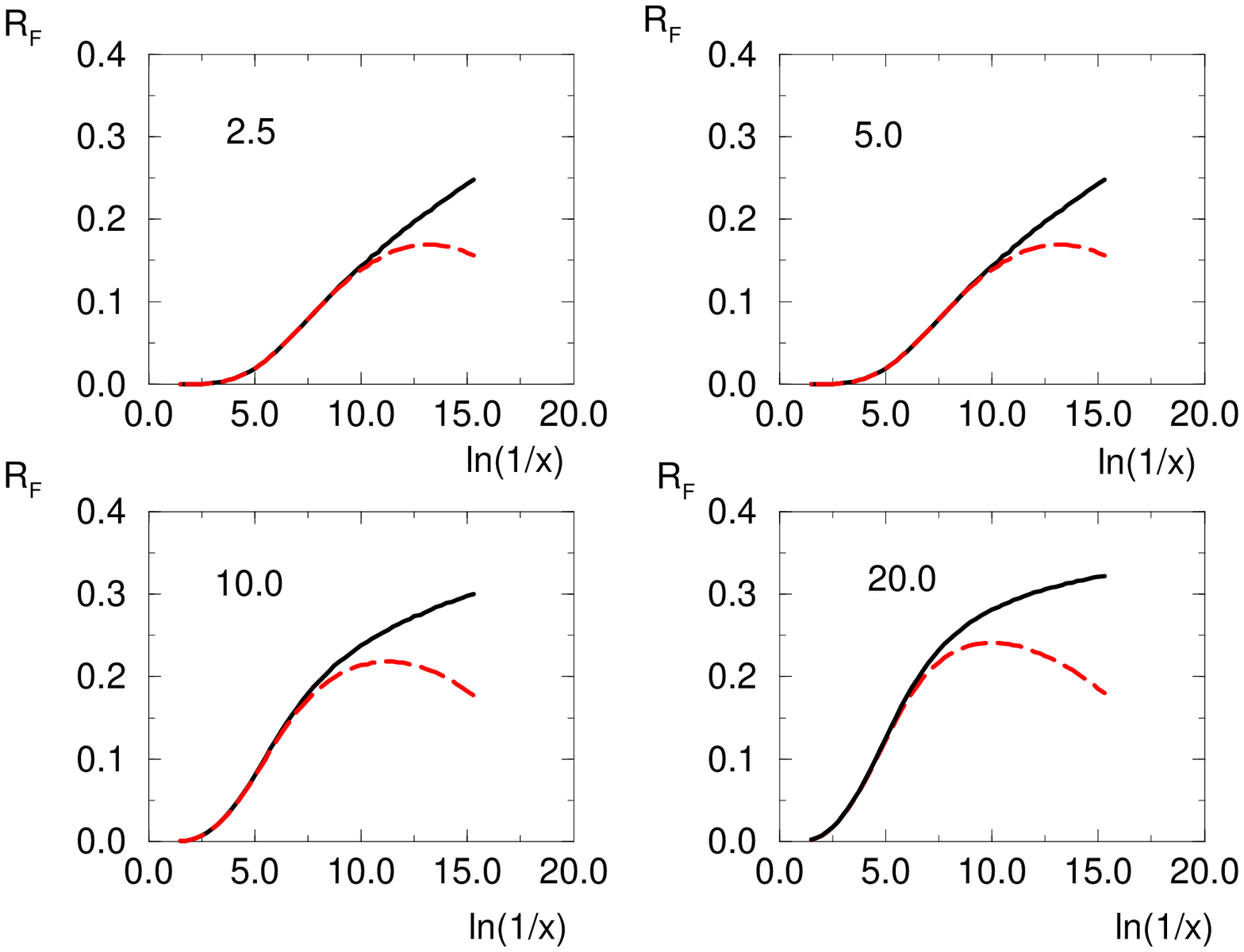,width=80mm}} 
\caption{  }
\label{fig79}
\end{figure}

This ratio is presented in Fig. (\ref{fig79}) \cite{45} as a function of $\ln
(1/x)$. There is strong modification of the ratio once UC are included in
the calculation. We urge data in this observable. The suppression is much
stronger than in the $F_2$ case, and we expect a lower production of quark
charm for small $x$, and this is related with the production of $J/\Psi$
which is proportional to the square of the gluon function.

Finally, one of our most stricking results concerns $eA$ physics, and is 
related with the high dense partonic system in the nuclear medium. The nuclear
shadowing is a challenge for hdQCD and mainly important for HERA-A, RHIC
and LHC physics.  We estimated how the nuclear structure
function and its derivative are modified by the  effects of high partonic
densitiy. 

\begin{figure}
\centerline{\psfig{file=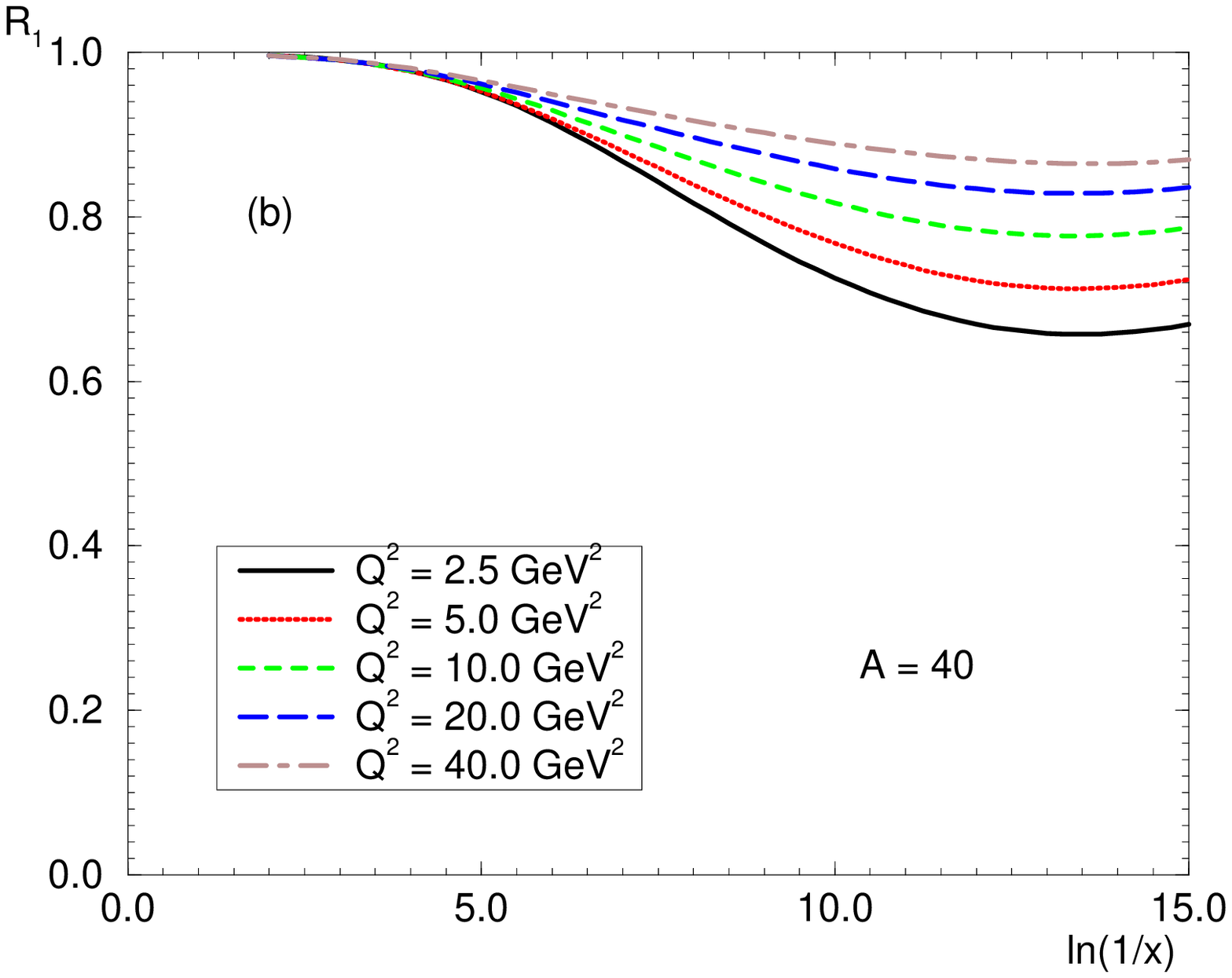,width=60mm}}
\caption{  }
\label{fig715}
\end{figure}

The shadowing corrections to $F_2^A$ are associated to the rescatterings of
the $q\bar{q}$ in the nucleons into the nucleus, being dependent on the
nucleon gluon distribution function. Here also we separate the two cases:
quark sector, where the gluon distribution is not modified by UC, and quark +
gluon sector, where now the gluon distribution is modified a la
Glauber-Mueller. The results are presented in Fig. (\ref{fig715}) \cite{51}
as a ratio $R_1=F_2^A/AF_2^N$, showing that for small $x$ the gluon sector
contribution should be included, and promote saturation.

We obtain that the  suppression due to the shadowing in $F_2^A$ is
proportionally smaller than in $xg_A$ in a perturbative framework, in a
different result that in \cite{r11} where soft physics is the main issue. As a
new result the saturation of the ratio is attained at HERA-A region when the
gluon sector is included. The presence of saturation in the perturbative
region ($Q^2 > 1$ GeV$^2$) denotes the large shadowing corrections in the
gluon sector.

The analysis was extended to the derivative of the nuclear structure function
\begin{eqnarray}
\frac{\partial F_2^A}{\partial \log Q^2} \frac{R^2_A Q^2}{2\pi^2} \sum_i e_i^2
\left[ C + \ln (\kappa_q) + E_1(\kappa_q) \right]\,,
\end{eqnarray}
considering the contributions of the quark and the gluon sectors to the UC.
for HERA-A $s=9.10^{4}$ GeV$^2$.

\begin{figure}
\centerline{\psfig{file=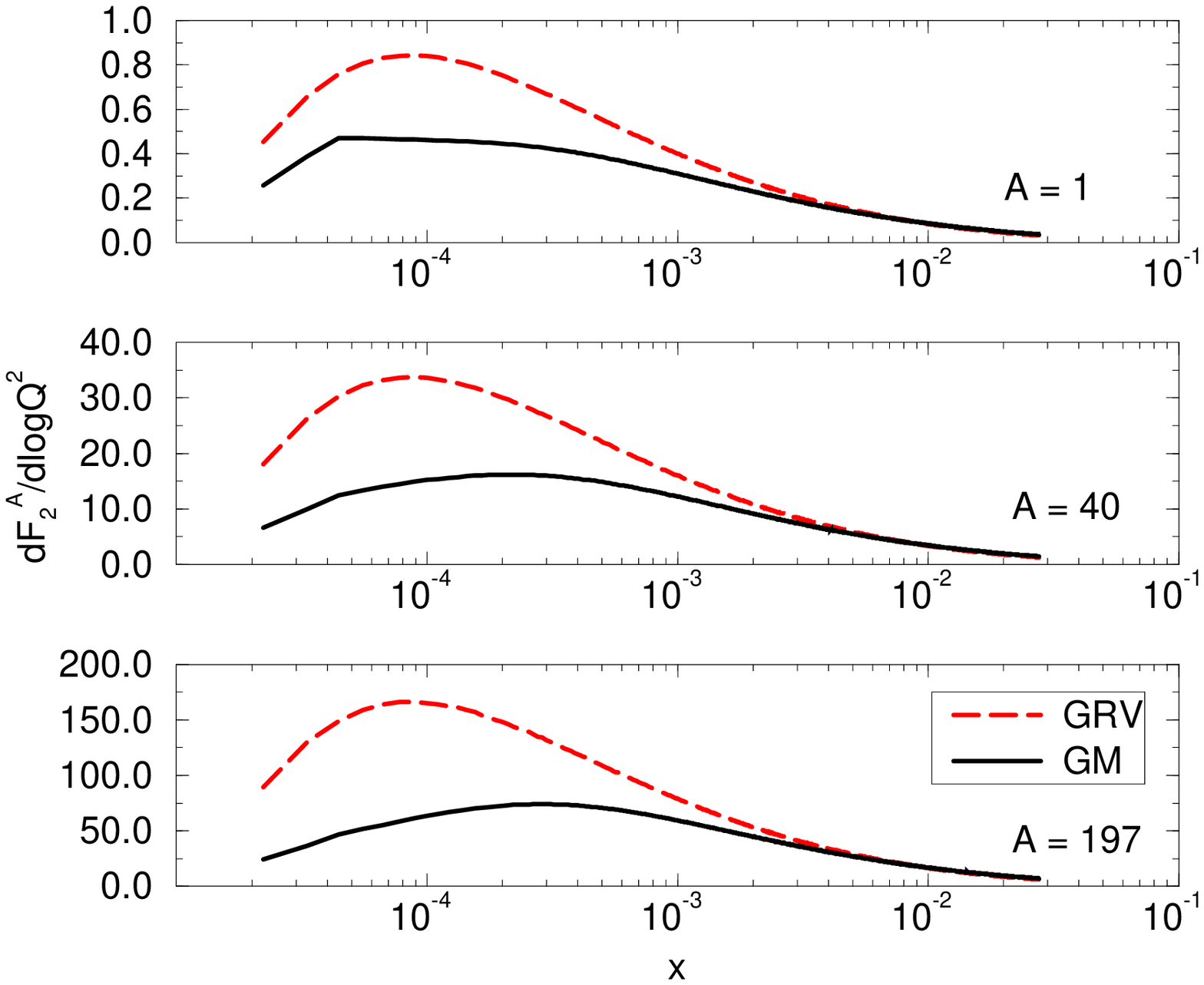,width=80mm}} 
\caption{  }
\label{fig718}
\end{figure}

The predictions are in Fig. (\ref{fig718}) \cite{52} compared with a DGLAP
calculation with GRV without nuclear effect. The expected turn-over is present
in the orthodox calculation but it is $A$ independent. The behavior of
the derivative  is different once UC are
considered since the maximum is $A$ dependent and runs to higher values of $x$
and $Q^2$ as $A$ increases. We conclude this is the best quantity to look for
unitarity corrections, evidentiating the same partonic density is reached as
$A$ increases for higher values of $x$ and higher values of $Q^2$,
corroborating a perturbative calculation. 

This is a strong motivation to develop this calculation for heavy
ion physics, and try to connect this formulation with the research in $AA$
physics where the quark-gluon plasma is expected to be produced.

\section{Outlook}

Several aspects of the formulations for hdQCD and in our approach to the
subject as well require further  investigation. I understand the formulation of
high dense partonic system should incorporate the methods of nonperturbative
physics and non-linear dynamics in order to present a comprehensive
formulation for a large kinematical regime in $ x$ and $Q^2$, besides
incorporating the $A$ dependence. However significative progress in the
description of hdQCD has been made in the recent years towards a unified
theoretical framework. Particularlly relevant is the role of initial
conditions for UC for the perturbative treatments and the determination of
saturation region, $Q_s^2$, still to be obtained analytically. Also a complete
solution of the generalized evolution equation (Eq. (26)) for $\alpha_s(Q^2)$
outside the asymptotic region is not available. Reaching these goals will
allow us to have a more complete dynamical description of the non-linear
phenomena of transition between large distance and short distance physics
promoting QCD to a more understandable and applicable theory.

\section*{Acknowledgments}

I thank the organizers of the XXI Encontro Nacional de F\'{\i}sica de
Part\'{\i}culas e Campos (ENFPC) for the kind  invitation  for this plenary
talk. The work presented here benefited of enlightening discussions with C. A.
Garcia Canal, F. Halzen and E. Levin in different phases of its developement.
I also thank my former students A. Ayala and V. Gon\c{c}alves for the lively
scientific atmosphere during the elaboration of their thesis, and my student
M. Machado for the invaluable criticism and help in the preparation of these
proceedings.

\end{document}